\documentclass[12pt]{article}
\usepackage[utf8]{inputenc}
\usepackage[T1]{fontenc}
\usepackage{lmodern}
\usepackage{setspace}
\usepackage{array}
\usepackage{amsthm} %command for \proof
\usepackage{amsmath}
\usepackage{caption}
\usepackage{graphicx}
\usepackage{float}
\usepackage{multirow}%%
\usepackage{color}
\usepackage{framed}
\usepackage{caption}
\usepackage{array}
\usepackage{amssymb}
\usepackage{amsmath}
\usepackage{subfigure}
\usepackage{booktabs,graphicx}%%
\usepackage{threeparttable}
\usepackage[dvipsnames]{xcolor}
\usepackage{color, soul}
\usepackage{soul}

\usepackage{natbib}
\setcitestyle{authoryear,round}
\usepackage{bm}
\usepackage{makecell}
\usepackage{algorithm,algorithmic}
\usepackage[colorlinks=true, citecolor=black]{hyperref}
\usepackage[margin=1in]{geometry}
\usepackage{url}

\newtheorem{theorem}{Theorem}
\newtheorem{lemma}{Lemma}

\definecolor{gray}{rgb}{0.9,0.9,0.9}
\definecolor{mygray}{rgb}{0.7,0.7,0.7}

\soulregister\cite7
\soulregister\citep7
\soulregister\citet7
\soulregister\ref7
\soulregister\pageref7
\soulregister\bm7

\definecolor{test}{RGB}{247,206,159} %247,206,159
\setulcolor{red}    %set underlining color  \ul{underlining}
\setstcolor{red}    %set overstriking color \st{overstriking}
\sethlcolor{test} %set highlighting color \hl{highlighting}  {yellow}Lavender, fill opacity=0.5 {Apricot}

\title{\vspace{-1.5cm}Independent policy gradient-based reinforcement learning for economic and reliable energy management of multi-microgrid systems}
\author{Junkai Hu\thanks{J. Hu is with the School of Mechanical and Electrical Engineering, Shenzhen Polytechnic University, Shenzhen
518055, China.}, \quad Li Xia\thanks{ L. Xia is with the School of Business, Sun Yat-Sen University, Guangzhou 510275, China. (email: xiali5@sysu.edu.cn)}}

\date{}

\begin{document}
\linespread{1.5}

\maketitle
\vspace{-0.2cm}
% \begin{frontmatter}

% \title{Independent policy gradient and reinforcement learning for reliable and economic energy management in multi-microgrid systems}

% \renewcommand{\thefootnote}{*}
% \author{\vspace{2mm}\normalsize Junkai Hu$^{a,}$\footnote{\textbf{Corresponding author}: Email address: @}
% \vspace{2mm}}

% \affiliation{organization={School of Business, Sun Yat-sen University, Guangzhou, 510275, P. R. China}}

\begin{abstract}
Efficiency and reliability are both crucial for energy management, especially in multi-microgrid systems (MMSs) integrating intermittent and distributed renewable energy sources.  This study investigates an economic and reliable energy management problem in MMSs under a distributed scheme, where each microgrid independently updates its energy management policy in a decentralized manner to optimize the long-term system performance collaboratively.
We introduce the mean and variance of the exchange power between the MMS and the main grid as indicators for the economic performance and reliability of the system. Accordingly, we formulate the energy management problem as a mean-variance team stochastic game (MV-TSG), where conventional methods based on the maximization of expected cumulative rewards are unsuitable for variance metrics. To solve MV-TSGs, we propose a fully distributed independent policy gradient algorithm, with rigorous convergence analysis, for scenarios with known model parameters. For large-scale scenarios with unknown model parameters, we further develop a deep reinforcement learning algorithm based on independent policy gradients, enabling data-driven policy optimization. Numerical experiments in two scenarios validate the effectiveness of the proposed methods. Our approaches fully leverage the distributed computational capabilities of MMSs and achieve a well-balanced trade-off between economic performance and operational reliability.

\end{abstract}


% \end{frontmatter}

%% \linenumbers

%% main text
\section{Introduction}

Microgrids function as the foundational components of smart grids, providing an operational environment for the efficient control and utilization of distributed energy resources and consumer demand loads. To mitigate the impact of uncertainties in these resources and demand loads, microgrids are typically equipped with energy storage devices, which can shift energy between different time intervals by discharging or charging, facilitating energy management \citep{weitzel2018energy}. However, the energy management capabilities of individual microgrids are often limited by their storage capacity. With advances in communication infrastructure, there is increasing attention on multi-microgrid systems (MMSs), where microgrids are interconnected through distribution buses and communication networks to better utilize distributed energy resources and mitigate power network uncertainties. Effective energy management methods can significantly improve both the economic efficiency and operational stability of MMSs \citep{alam2018networked}.

\begin{figure}[!htbp] 
    \centering \includegraphics[width=10cm]{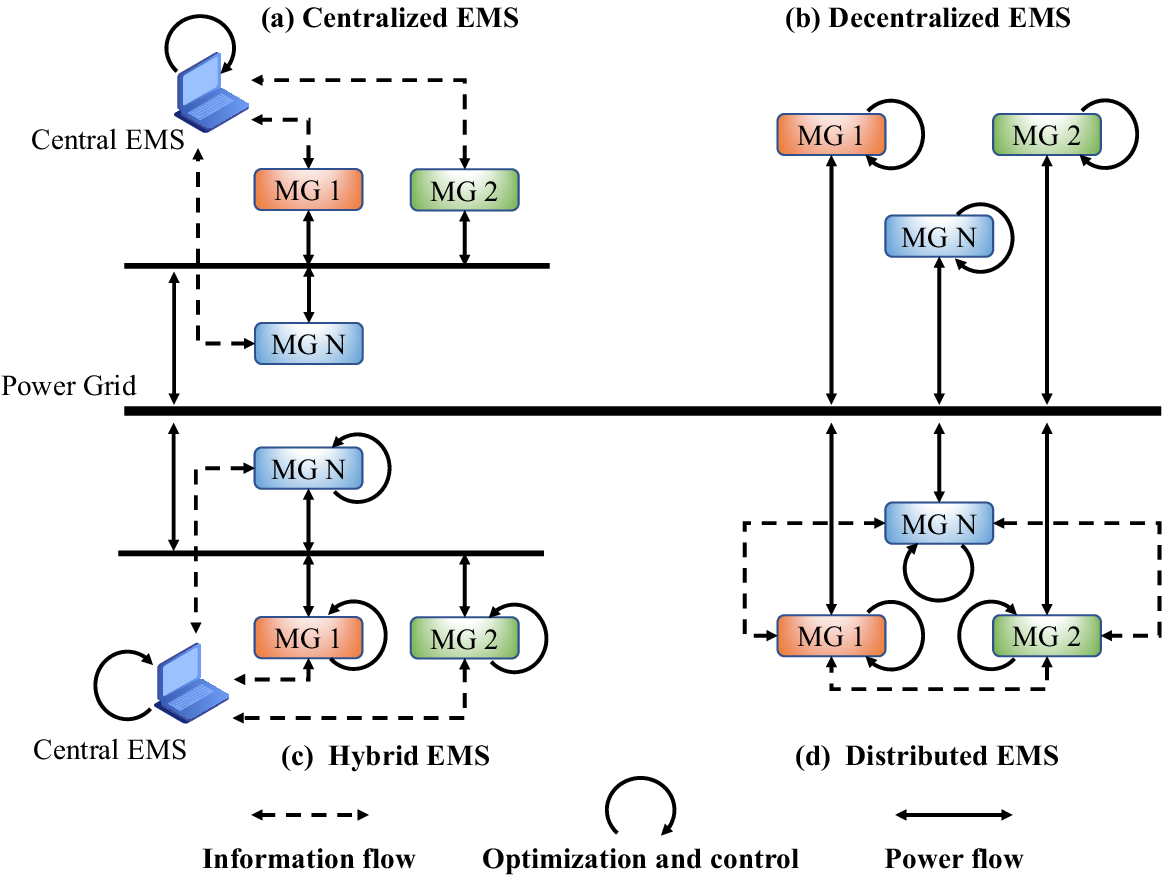} \caption{The illustration of  EMS topologies in MMSs (MG denotes the microgrid).} 
    \label{figure:ems_topo} 
\end{figure}

In MMSs, each microgrid is equipped with its own energy management system (EMS) and implements a local policy for controlling its dispatchable resources. To facilitate collaboration among microgrids, various EMS topologies have been proposed, generally classified into four categories  \citep{nawaz2022comprehensive}, as illustrated in Figure~\ref{figure:ems_topo}: Centralized EMS; Decentralized EMS; Hybrid EMS; Distributed EMS. The distributed scheme stands out for the absence of a centralized controller or coordinator. Instead, each microgrid communicates with neighboring microgrids via communication infrastructure and updates its energy management policy accordingly. This approach leverages the distributed computational capabilities of MMSs and avoids a single point of failure. In this work, we focus on the energy management problem in MMSs under the distributed EMS scheme.

A majority of the literature on energy management in MMSs focuses on achieving economic objectives within rigid operational constraints, such as minimizing operational costs \citep{chen2022research} and increasing the penetration of renewable energy \citep{han2024optimal}. However, integrating a large number of uncertain units can negatively impact the reliability and stability of power network operations. The intermittent and unpredictable nature of renewable energy generation leads to significant power fluctuations, which may cause various power quality issues, including voltage instability and flicker effects \citep{yang2020fluctuation}.

To mitigate the fluctuations in exchange power, a thread of work focuses on power smoothing by formulating optimization objectives based on power ramp rates, which quantify the rate of power change between two consecutive time intervals \citep{arevalo2023smart,wang2023coordinated}. Another thread of work aims to achieve peak shaving and valley filling within the scheduling horizon, which is also a widely adopted practical objective \citep{tang2021hierarchical, manojkumar2022rule}. It is evident that in  MMS energy management, where renewable energy generation and loads are uncertain, reducing exchange power fluctuations is essential to ensuring the reliable operation of the overall system.

Traditional energy management methods, such as programming-based approaches, typically rely on accurate system modeling or precise predictions of uncertain variables. However, the increasing integration of stochastic components and the expanding scale of power networks have substantially heightened system complexity, posing significant challenges to conventional model-based optimization and programming techniques. In response, data-driven approaches have emerged as a promising direction and have attracted widespread attention in the field of energy management \citep{li2023methods}.

Reinforcement learning (RL) has recently gained growing interest in energy management for power networks, due to its notable success in complex decision-making tasks, ranging from mastering the game of Go \citep{silver2017mastering} to empowering large language models. RL is a data-driven approach for solving stochastic sequential decision-making problems, typically modeled as Markov decision processes (MDPs). It enables an agent to repeatedly interact with an environment with unknown dynamics and rewards, collecting data on state transitions and rewards to iteratively update its policy and maximize expected cumulative rewards. Among various RL approaches, policy gradient methods are the most widely used, where the policy is iteratively updated along the gradient direction of the objective performance function \citep{agarwal2021theory}. Additionally, compared to traditional tabular RL methods, deep reinforcement learning (DRL) incorporates deep neural networks to approximate value and policy functions, alleviating the curse of dimensionality and enhancing generalization and scalability.

However, standard RL or DRL methods are primarily designed to maximize the expected cumulative rewards. In many real-world applications, particularly in power networks, it is essential to consider reliability and risk-sensitive performance metrics, which are usually variance-related and do not exhibit the Markovian or additive characteristics of reward functions. These features make the dynamic programming principle fail \citep{filar1989variance,xia2020risk} and limit the direct applicability of conventional RL algorithms to economic and reliability-aware energy management problems.

In this work, we investigate long-term economic and reliable energy management in MMSs under the distributed EMS scheme, aiming at updating the energy management policies of individual microgrids to optimize the exchange power between the MMS and the main grid. The average exchange power over the scheduling horizon corresponds to electricity sold (positive) or purchased (negative), reflecting the economic aspect of MMS operation. We introduce the variance metric as an additional optimization objective to capture the volatility of exchange power, which reflects the reliability and stability of MMS operation. Since the problem requires multiple microgrids to jointly optimize a variance-based metric, standard policy gradient and RL methods, which only maximize the expected cumulative rewards, are not directly applicable. To address this challenge, we formulate the problem as a mean-variance team stochastic game (MV-TSG) and propose corresponding algorithms to solve it. The main contributions of this paper are as follows:
\begin{itemize}
    \item Based on the characteristics of the distributed EMS scheme, we propose a mean-variance independent projected gradient ascent (MV-IPGA) method for MV-TSGs, as illustrated in Algorithm~\ref{alg:independent_pg}. In this approach, each microgrid independently and simultaneously updates its energy management policy to improve the overall economic and reliability performance of MMS operation. We also establish the global convergence of the proposed algorithm. Compared to methods that require centralized coordination, the proposed approach utilizes the distributed computing resources in MMSs more efficiently.
    \item Building upon Algorithm~\ref{alg:independent_pg}, we further develop an independent policy gradient-based reinforcement learning algorithm, referred to as mean-variance independent proximal policy optimization (MV-IPPO), as presented in Algorithm~\ref{alg:independent_ppo}. This algorithm enables approximate solutions to large-scale MV-TSGs in a data-driven manner when environmental parameters are unknown.
\end{itemize}
The effectiveness of the proposed algorithms is validated through field-data-based experiments under two distinct scenarios: one with fully known model parameters and another with unknown parameters. The results demonstrate that our methods successfully achieve a balance between the operational economy and reliability of MMSs within a distributed EMS framework. Compared with existing works \cite{qiu2021rmix} and \cite{shen2023riskq}, this study, to the best of our knowledge, is the first to develop a \textit{fully distributed, risk-sensitive, cooperative RL algorithm with theoretical guarantees in heterogeneous multi-agent settings.}

The remainder of this paper is organized as follows. 
Section~\ref{sec:literature review} reviews the relevant literature on reducing power fluctuations in microgrids and risk-sensitive RL. 
Section~\ref{sec:problem formulation} presents the problem formulation and the MV-TSG model. The MV-IPGA method and its convergence analysis are introduced in Section~\ref{sec:MV-IPGA}. The MV-IPPO algorithm is further proposed in Section~\ref{sec:MV-IPPO}.
Section~\ref{sec:experiments} provides the numerical experiments, and Section~\ref{sec:conclusion} concludes the paper.

\section{Literature review}
\label{sec:literature review}
\noindent In this section, we first review the literature on reducing power fluctuations in microgrids. Subsequently, we briefly introduce the risk-sensitive RL, which optimizes some specific risk metrics instead of the expected cumulative reward. 

\subsection{Reducing power fluctuations in microgrids}
Reducing power fluctuations is essential for the reliable operation of power networks. Some studies focus on power smoothing by optimizing objectives related to power ramp rates. For example, \cite{arevalo2023smart} investigate photovoltaic power smoothing based on failure detection. They propose a method combining moving averages and ramp rate control, with energy storage systems (ESSs) to mitigate power fluctuations, and
validate its effectiveness in a microgrid laboratory at the University of Cuenca. Similarly, \cite{abdalla2023novel} study power smoothing by incorporating cloud information in a two-layer framework: the first layer predicts and classifies cloud types, while the second dynamically adjusts the filter time constant based on power ramp rates. However, the effectiveness of such approaches heavily depends on the accuracy of the predictions, which becomes increasingly difficult as the time horizon extends. \cite{wang2023coordinated} address power smoothing for multiple wind turbines with energy storage using a multi-agent reinforcement learning (MARL) algorithm. They reshape the reward function by incorporating the power ramp rate. Nonetheless, the aforementioned approaches, which optimize power ramp rates, are limited to mitigating fluctuations between two successive time intervals, and are insufficient for achieving peak shaving and valley filling over the entire scheduling horizon.

Peak shaving is another critical objective in energy management for reducing power fluctuations. \cite{guo2021cooperative} study a cooperation problem involving distributed generators, ESSs, and voltage regulating devices, aiming to minimize daily operational costs while constraining peak load demand. They propose a two-stage programming method, with the peak demand limit determined via trial and error. \cite{ghafoori2023electricity} focus on optimal scheduling for electric vehicle charging and discharging to minimize peak power demand in commercial buildings. Their approach reduces the variance between actual and forecasted minimum demand across all time intervals, using machine learning for demand forecasting and binary linear programming for optimization. Similarly, \cite{chapaloglou2019smart} propose a predict-then-optimize framework for load smoothing and peak shaving in microgrids. However, these methods are primarily designed for short-term, intra-day scheduling and are less applicable to long-term energy management due to the growing difficulty of accurate predictions over extended time horizons.

Markov model is a powerful tool for characterizing stochastic dynamic systems. \cite{yang2020fluctuation} investigate long-term energy management for microgrids to reduce fluctuations in the exchange power with the main grid. They formulate the problem as an MDP with a variance objective and propose a policy iteration type method to solve it, effectively addressing both power smoothing and peak shaving in long-term scheduling.  \cite{peirelinck2024combined} and \cite{rostmnezhad2023power}  investigate demand response optimization and building energy management problems, respectively, both aiming to reduce peak power. They formulate these problems as MDPs and solve them using DRL approaches. \cite{hu2024economical} study an MMS energy management problem focusing on economic and reliable long-term operation. They incorporate a variance objective to reflect reliability and propose a MARL algorithm based on a centralized training with decentralized execution (CTDE) framework. However, all aforementioned studies rely on centralized control or coordination, which limits their applicability to the distributed MMS energy management problem studied in this work.

\subsection{Risk-sensitive reinforcement learning}
RL is a subfield of machine learning that has been successfully applied across various domains to address complex sequential decision-making problems, including energy management in power networks. Policy gradient methods, such as proximal policy optimization (PPO) \citep{schulman2017proximal}, are the most popular type of RL algorithms, which utilize the gradient information of the performance function to guide policy updates. However, in power networks, the complexity and inherent uncertainty of real-world scenarios suggest that maximizing cumulative rewards (e.g., economic gains) alone is insufficient.  To ensure practical and safety-critical operations, it is crucial to incorporate risk and system reliability metrics into the decision-making process \citep{blancas2024discrete}.

Risk-sensitive RL is a subfield of safe RL and has long been a prominent research direction \citep{garcia2015comprehensive}. It focuses on optimizing performance criteria that account for specific risks. Common risk metrics include variance, conditional value-at-risk (CVaR), absolute deviation, and semi-variance. However, many of these measures lack additivity and may not satisfy the Markovian property, which causes the Bellman equation—the foundation of standard RL methods—to no longer hold  \citep{bauerle2024markov,ma2023unified}. Consequently, conventional RL algorithms often struggle with optimization problems involving such risk metrics, requiring the development of specialized approaches to address these challenges.

In the context of variance optimization in RL or MDPs,  existing literature can be classified into two groups based on the variance definition. The first group concerns the variance of cumulative rewards, i.e., $\text{Var}(\sum\limits_{t=0}^{\infty} \gamma^t r_t)$, where $\gamma$ is the discount factor and $r_t$ is the feedback reward at time step $t$ \citep{prashanth2022risk,huang2018finite}. This formulation is typically used to quantify fluctuations in total profits or costs.  The second group focuses on the long-run variance or steady-state variance \citep{xia2018variance,bisi2021risk,filar1989variance}, defined as $\lim\limits_{T \to \infty}\frac{1}{T} \mathbb{E}_{\mu} [\sum\limits_{t=0}^{T-1}(r(s_t,a_t)-\eta^{\mu})^2]$, where $\eta^{\mu}$ is the long-run average reward under policy $\mu$. This metric captures long-term reward volatility and can serve as an indicator of system operational stability.

Due to the non-additivity of variance, standard RL algorithms are generally inadequate for solving either type of variance-based optimization problem. As a result, specialized policy gradient algorithms are often designed to address these challenges \citep{prashanth2022risk}. However, in the energy management problem of MMSs under the distributed EMS scheme, each microgrid operates independently, making decisions and updating its policy in a decentralized manner. When single-agent RL algorithms are used independently by each microgrid, and policies are updated simultaneously, the environment perceived by each microgrid becomes non-stationary.  This violates the stationary environment assumption of single-agent RL, and leads to a lack of guarantees for improvement in the common objective function \citep{zhong2024heterogeneous}. Therefore, directly applying single-agent RL methods in this setting is usually inappropriate, which highlights the complexity and challenges of cooperative risk-sensitive optimization in multi-agent settings.

A related research direction is cooperative MARL, which aims to learn behavior policies for multiple independent agents that collectively optimize a shared objective. Given the prevalence of cooperative systems in real-world applications \citep{rosa2024modular}, this remains an active research area in MARL. However, few studies have addressed the collective optimization of risk-sensitive metrics. \cite{qiu2021rmix} and \cite{shen2023riskq} investigate cooperative MARL with objective functions such as CVaR and weighted quantile. Their approaches focus on the design of neural network architectures to approximately satisfy certain optimization conditions, thus lacking theoretical analysis. Moreover, both of their algorithms follow the CTDE framework, which requires centralized coordination during the training phase and suffers from high computational burdens \citep{chen2025scalable}. To the best of our knowledge, no effective algorithms with theoretical guarantees exist for fully distributed cooperative MARL under risk-sensitive objectives.

\section{Problem description and modeling}
\label{sec:problem formulation}
In this section, we first briefly introduce the MMS under the distributed EMS scheme. Subsequently, we model the economic and reliable energy management of MMSs as an MV-TSG, and provide details for variable definitions and corresponding constraints.
\subsection{Multi-microgrid system}
\begin{figure}[!htbp] 
    \centering \includegraphics[width=8cm]{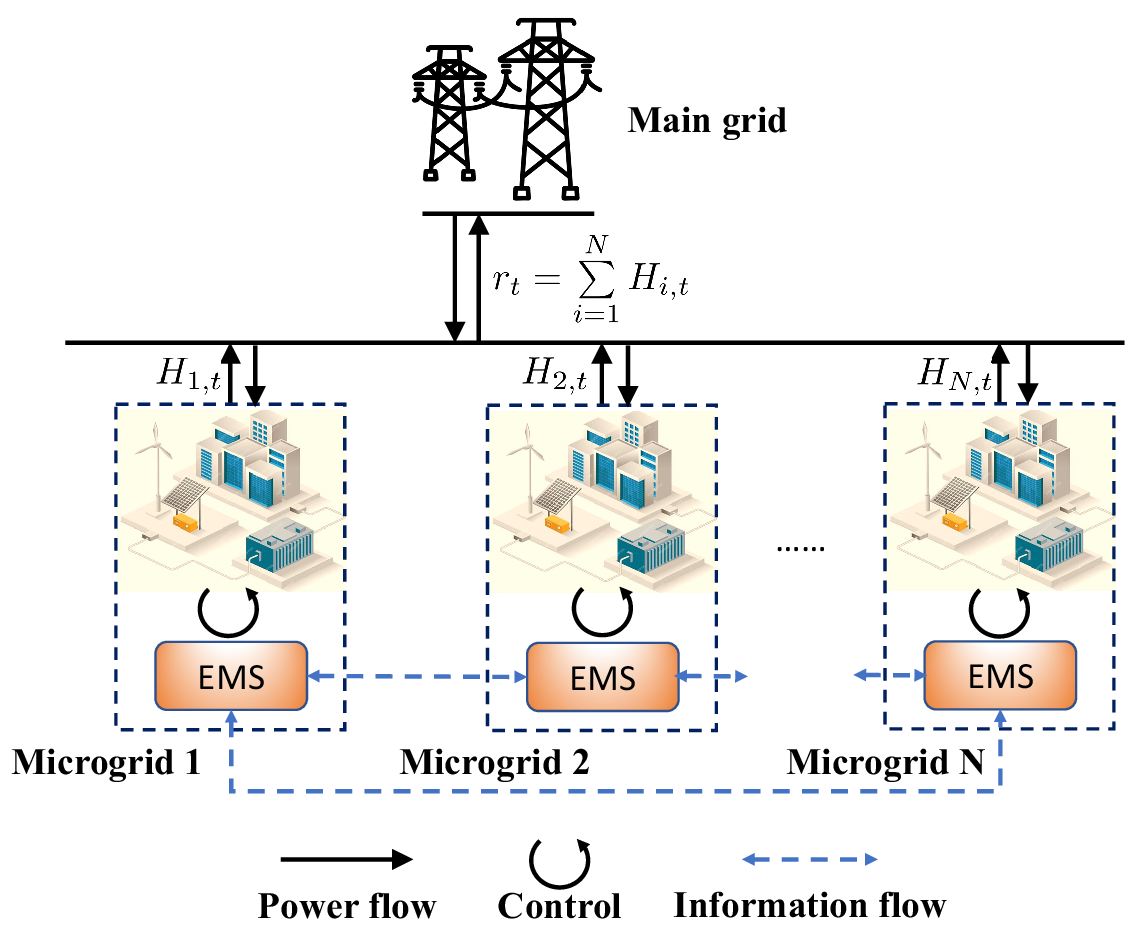} \caption{Grid-connected MMS under the distributed EMS scheme} 
    \label{figure:distribued-ems} 
\end{figure}

In this study, we investigate an MMS under the distributed EMS scheme, as illustrated in Figure~\ref{figure:distribued-ems}. The MMS consists of a set $\mathcal{N} = \{1, 2, \dots, N\}$ of $N$ microgrids. Each microgrid may include renewable energy generators, demand loads, and ESSs. The renewable energy generator, such as a wind turbine or a solar panel, generates power by harvesting energy from renewable resources. Demand loads represent power consumers whose consumption cannot be curtailed. Consequently, microgrids are both energy producers and consumers, with bidirectional power flow. The ESS facilitates energy management by shifting energy across time steps through real-time charging or discharging. For example, when renewable energy generation exceeds the demand load, the ESS stores the surplus energy; conversely, when the demand load surpasses generation, the ESS discharges stored energy to meet the power shortfall. In the distributed EMS scheme, each microgrid operates an individual EMS and can communicate with neighboring microgrids through the communication infrastructure.

For microgrid $i \in \mathcal{N}$, the maximum output power of the renewable energy generator is denoted by $G_i^{\text{max}}$, and the minimum and maximum consumed power of the demand load is denoted by  $D_i^{\text{min}}$ and $D_i^{\text{max}}$, respectively.
At time step $t=1,2,\dots$, the renewable energy generated power and demand load of microgrid $i$ are denoted by $G_{i,t}$ and $D_{i,t}$, respectively. 
Since both depend on random factors such as climate conditions, $G_{i,t}$ and $D_{i,t}$ are non-negative random variables. Each microgrid $i$ is also equipped with an ESS subject to minimum and maximum energy level constraints, denoted by $B_i^{\text{min}}$ and $B_i^{\text{max}}$, respectively. At time step $t$, the energy level of the ESS is denoted by $B_{i,t}$, and the ESS is discharging with power $b_{i,t}$ (where $b_{i,t}<0$ indicates charging). Due to power loss during the charging and discharging processes, we introduce an efficiency coefficient $\nu$ for both operations. The actual output or absorbed power of the ESS is given by $e_{i,t} = \nu [b_{i,t}]^{+} + \frac{1}{\nu} [b_{i,t}]^{-}$, where $[b_{i,t}]^{+} = \max \{0,b_{i,t}\}$ and $[b_{i,t}]^{-} = \min\{0,b_{i,t}\}$.

We define the discrete sets for the generated power, demand load power, and storage energy level of microgrid $i$ as $\mathcal{G}_i := \{0, \dots, G_i^{\text{max}}\}$, $\mathcal{D}_i := \{D_i^{\text{min}}, \dots, D_i^{\text{max}}\}$, and $\mathcal{B}_i := \{B_i^{\text{min}}, \dots, B_i^{\text{max}}\}$, respectively. For microgrids with renewable energy generators, power curtailment is permitted when necessary. The EMS of each microgrid controls the ESS discharging power $b_{i,t}$ and the curtailed power $v_{i,t}$ from the renewable energy generator. The output power of microgrid $i$ is then given by $H_{i,t}=G_{i,t}-D_{i,t}+e_{i,t}-v_{i,t}$.  

Consequently, the total output power of the MMS or the exchange power between the MMS and the main grid is $H_t={\sum\limits_{i=1}^N H_{i,t}}$. When $H_t>0$, the MMS is outputting power to the main grid and getting profits. Conversely, when $H_t<0$, the MMS purchases power from the main grid to meet demand or charge the ESSs, incurring additional costs. Moreover, fluctuations in exchange power may lead to power quality issues and serve as a reliable indicator for MMS operations.

In this work, we consider the economic and reliable energy management problem of MMSs, and use the variance to measure the fluctuations of exchange power. Given that the MMS consists of multiple microgrids with individual energy management policies, we model the MMS energy management problem as a team stochastic game (TSG) due to its advantages in dealing with stochastic dynamic systems with multiple decision makers.  

\subsection{Mean-variance team stochastic game model}
% 写的是一般的形式
A long-run TSG is defined by a tuple $\langle \mathcal{N}, \mathcal{S},\mathcal{A}, P, r\rangle$. Here, $\mathcal{N}=\{1,\dots, N\}$ is a finite set of agents, $\mathcal{S}$ is a finite set of system states, $\mathcal{A}$ is the joint action space of all agents, $\mathcal{A}_i$ is the action space of agent $i\in \mathcal{N}$,  $P:\mathcal{S} \times \mathcal{A} \mapsto \Delta(\mathcal{S})$ is the transition probability function ($\Delta(\mathcal{S})$ is the probability distributions over $\mathcal{S}$), $r:\mathcal{S} \times \mathcal{A} \mapsto \mathbb{R}$ is a bounded common reward function. In the long-run TSG, the stationary policy for each agent $i \in \mathcal{N}$ is a mapping $\mu_i:\mathcal{S}\mapsto \Delta(\mathcal{A}_i)$ and the corresponding stochastic policy space is denoted as $\mathcal{U}_i$. The joint policy is denoted by $\bm\mu=(\mu_1,\dots,\mu_N)$ and its corresponding policy space is $\mathcal{U}=\prod\limits_{i=1}^N\mathcal{U}_i$. 

The dynamic decision-making process of multiple agents is stated as follows. At each time step $t$, each agent selects an action $a_{i,t} \in \mathcal{A}_i$ according to the system state $s_t$ and its randomized policy function $\mu_i(\cdot|s_t)$. Subsequently, the joint action $\bm{a_t}=(a_{1,t},\dots, a_{N,t})$ lead to a common reward $r_{t+1}=r(s_t,\bm{a}_t)$ to all agents and induces the system's transition to a next state $s_{t+1}$ according to the state transition function $P(s_{t+1}|s_t,\bm{a}_t)$. The entire decision-making process generates a trajectory $\tau=\{s_0,\bm{a}_0,r_1,s_1,\bm{a}_1,r_2,s_2,\dots \}$.

The variable definitions for the TSG in the context of specific MMS energy management problems are described below.

\begin{itemize}
    \item Number of agents $N$: Each microgrid corresponds to an agent, and $N$ denotes the total number of microgrids in the MMS. 
    \item System state $s_t$: In MMSs under the distributed EMS scheme, each microgrid can communicate with each other, and the state variables of MMSs are fully observable for each microgrid. The MMS state should include the state variables of all microgrids, including the output power of the renewable energy generator, the demand load power, and the storage energy level. Then, the system state is given as $s_t=\{G_{1,t},D_{1,t},B_{1,t},\dots, G_{N,t},D_{N,t},B_{N,t}\}$.
    \item Action $a_{i,t}$: At each time step, the EMS of each microgrid $i$ can conduct energy management by controlling the output power $b_{i,t}$ of ESSs and the curtailed power $v_{i,t}$ of renewable energy generators. Then, the action is given as $a_{i,t}=(b_{i,t},v_{i,t})$. 
    \item State transition $P$: We assume that the variables $G_{i,t}$ and $D_{i,t}$ of each microgrid $i$ follow independent and identically distributed random processes, which can be described by Markov chains. This assumption generally holds when the microgrids are located relatively far from each other \citep{etesami2018stochastic}. The state transition of ESSs is given by $B_{i,t+1}=B_{i,t}-b_{i,t}$, where the next storage energy level depends on the action taken at the current time step. Then, the state transition $P$ of the MMS is determined by the randomness in renewable energy generation and demand loads, as well as the behavior policy.
    \item Reward $r(s_t,\bm{a}_t)$: In this study, we define the common reward as the exchange power between the MMS and the main grid, i.e., $r(s_t,\bm{a}_t)=H_t=\sum_{i=1}^N (G_{i,t}-D_{i,t}+e_{i,t}-v_{i,t})$. A positive reward,  $r(s_t,\bm{a}_t)>0$, indicates that the MMS is supplying power to the main grid and generating profit.
\end{itemize}

During the MMS operation, certain constraints must be satisfied in the dynamic operation of MMSs. Specifically, for $t=1,2,\dots$, the ESS should meet the capacity constraints: $B_i^{\text{min}}\le B_{i,t} \le B_i^{\text{max}}$ and $B_{i,t}-B_i^{\text{max}}\le b_{i,t} \le B_{i,t}-B_i^{\text{min}}$. Moreover, the storage output power should also satisfy the power constraint of the interface inverters of the ESS: $-C_i^{\text{ch}}\le b_{i,t} \le C_i^{\text{dis}}$, where $C_i^{\text{ch}}$ and $C_i^{\text{dis}}$ denote the maximum charging and discharging power, respectively. Besides, the curtailed power of renewable energy generators should be no more than the power that can be generated, $0 \le v_{i,t} \le G_{i,t}$.

In this study, we consider the economic and reliable energy management problem of
MMSs. For the economic aspect, the long-run average reward is regarded as the objective function,
\begin{equation}
        \eta^{\bm{\mu}}:=\lim\limits_{T \rightarrow \infty} \frac{1}{T} \mathbb{E}_{\bm{\mu}} \Big[ \sum\limits_{t=0}^{T-1}r(s_t,\bm{a}_t)  \Big], \notag
\end{equation}
which reflects the electricity sales revenue or procurement cost during the MMS long-term operation.
With $\pi^{\bm{\mu}}$ denoting the steady state distribution under the joint policy $\bm{\mu}$, it is convenient to rephrase the long-run average reward as
\begin{align}
    \eta^{\bm{\mu}}
    &= \mathbb{E}_{s\sim \pi^{\bm{\mu}}, \bm{a} \sim \bm{\mu}}[r(s,\bm{a})] \notag\\
    &= \sum\limits_{s \in \mathcal{S}} \pi^{\bm\mu}(s) \sum\limits_{\bm{a} \in \mathcal{A}} r(s,\bm{a})\bm\mu(\bm{a}|s).  
    \label{eq:eta}
\end{align}
For the system operational reliability, we introduce the long-run variance of exchange power, defined as
\begin{equation}\label{eq:zeta}
    \zeta^{\bm{\mu}} := \lim\limits_{T \rightarrow \infty} \frac{1}{T} \mathbb{E}_{\bm{\mu}} \Big[ \sum\limits_{t=0}^{T-1}(r(s_t,\bm{a}_t) - \eta^{\bm{\mu}} )^2   \Big], \notag
\end{equation}
which describes the fluctuation of exchange power in a long-term perspective. Since a power grid needs to maintain a strict balance between power supply and demand, large fluctuations in exchange power will degrade the stability and safety of the power system.

Then, the MV-TSG aims to maximize the expected mean-variance combined metric
\begin{align}\label{eq:mean-variance}
     J^{\bm{\mu}} 
     &= \lim\limits_{T \rightarrow \infty} \frac{1}{T} \mathbb{E}_{\bm{\mu}} \Big\{ \sum\limits_{t=0}^{T-1} [r(s_t,\bm{a}_t) - \beta(r(s_t,\bm{a}_t) - \eta^{\bm{\mu}} )^2]   \Big\} \\
     &=  \eta^{\bm{\mu}} - \beta \zeta^{\bm{\mu}}, \notag
\end{align}
where $\beta \ge 0$ is the parameter for the trade-off between mean and variance. Next, we propose an independent policy gradient method and a DRL approach to address MV-TSGs.

\section{Independent policy gradient}
\label{sec:MV-IPGA}
In this section, we first present the value function definitions and corresponding policy gradient results for MV-TSGs. Subsequently, we give the exact policy gradient method for MV-TSGs and analyze its global convergence properties.

\subsection{Value functions and properties}
In this work, we investigate the independent policy gradient method in the energy management problem. We consider direct decentralized policy parameterization, where each agent's policy is parameterized by $\theta_i$,
\begin{equation*}
    \mu_{i}^{\theta_i}(a_i|s)=\theta_{i,s,a_i}, \quad i=1,2,\dots,N.
\end{equation*}
For any $s\in \mathcal{S}$ and $a_i$, we have $\theta_{i,s,a_i}>0$ and $\sum\limits_{a_i \in \mathcal{A}_i} \theta_{i,s,a_i}=1$. For notational simplicity, we abbreviate $\mu_{i}^{\theta_i}(a_i|s)$ as $\mu^{\theta_i}(a_i|s)$, $\theta_{i,s,a_i}$ as $\theta_{s,a_i}$. The joint policy $\bm\mu^{\theta}(\bm{a}|s)=\prod\limits_{i=1}^N \mu^{\theta_i}(a_i|s)=\prod\limits_{i=1}^N \theta_{s,a_i}$. We use $\mathcal{X}_i:=\Delta(\mathcal{A}_i)^{|S|}$, $\mathcal{X}:=\mathcal{X}_1\times \cdots \times \mathcal{X}_N$ denote the feasible region of $\theta_i$ and $\theta=(\theta_1,\dots,\theta_N)$, respectively. In the rest of the paper, we use the notation $-i$ to denote the set of all agents except agent $i$.

Inspired by Equation~(\ref{eq:mean-variance}), we provide a surrogate reward function as 
\begin{equation}
    f^{\theta}(s,\bm{a})=r(s,\bm{a}) - \beta(r(s,\bm{a})-\eta^{\theta})^2.
    \label{eq:para_reward_f}
\end{equation} 
Subsequently, similar to Equation~(\ref{eq:eta}), the mean-variance performance can be calculated by
\begin{equation*}
    J^{\theta}=\sum\limits_{s \in \mathcal{S}} \pi^{\theta}(s) \sum\limits_{\bm{a} \in \mathcal{A}} f^{\theta}(s,\bm{a})\bm\mu^\theta(\bm{a}|s).
    % \label{eq:para_eta_f}
\end{equation*}

Following the definitions of average-reward MDPs in \cite{sutton2018reinforcement}, the value function $V_{f}^{\theta}$, action-value function $Q_{f}^{\theta}$ and advantage function $A_{f}^{\theta}$, with respective to the joint policy and surrogate reward function $f^{\theta}$, are defined as 
\begin{align*} 
    % \label{eq:para_Vf function} 
    V_{f}^{\theta}(s)
    :&= \mathbb{E}_{\bm\mu^\theta} \Big[\sum\limits_{t=0}^\infty (f^{\theta}(s_t,\bm{a}_t)- J^{\theta})|s_0 = s \Big], \\
    % \label{eq:para_qf function} 
    Q_{f}^{\theta}(s,\bm{a}):&=\mathbb{E}_{\bm\mu^\theta} \Bigl[
        \sum\limits_{t=0}^\infty 
        (f^{\theta}(s_t,\bm{a}_t) -J^{\theta})|s_0=s,\bm{a}_0=\bm{a}
    \Bigr]  \\
    & = f^{\theta}(s,\bm{a})  -J^{\theta} + \sum\limits_{s' 
    \in \mathcal{S}}P(s'|s,\bm{a})V_{f}^{\theta}(s'),  
    % \label{eq:para_qf function-2 }
    \\
    % \label{eq:para_af function} 
    A_{f}^{\theta}(s,\bm{a}):&= Q_{f}^{\theta}(s,\bm{a}) - V_{f}^{\theta}(s).
\end{align*}

We now present the performance difference lemma in MV-TSGs, which quantifies the mean-variance performance difference between any two joint policies. Lemma~\ref{lemma:para_perf_diff} is a direct extension of the mean-variance difference formula for MDPs, as presented in \cite{xia2020risk}, to the TSG setting, with the proof omitted for brevity. For clarity, the notations $J(\bm\mu),J^{\bm\mu},J(\theta)$ and $J^{\theta}$ are used interchangeably when necessary.

\begin{lemma}[Performance difference in MV-TSGs]
    For any two joint policies $\bm\mu, \bm\mu' \in \mathcal{U}$, the mean-variance performance difference is
   \begin{align*}
        % \label{eq:MVPDF-1
        J(\bm{\mu}') - J(\bm{\mu}) 
        = \mathbb{E}_{s \sim \pi^{\bm{\mu}'}, \bm{a} \sim \bm{\mu}'}[A_{f}^{\bm{\mu}}(s,\bm{a})] + \beta(\eta^{\bm\mu'} - \eta^{\bm\mu} )^2.
    \end{align*}
    \label{lemma:para_perf_diff}
\end{lemma}

Different from the standard discounted or averaged reward cases, where the reward function $r$ is independent of policy $\mu$, the surrogate reward function $f$ defined in (\ref{eq:para_reward_f}) includes a term of $\eta$, which depends on the joint policy. Then, a new policy gradient lemma is derived for MV-TSGs based on Lemma~\ref{lemma:para_perf_diff} and stated as follows.
\begin{lemma}[Mean-variance policy gradient]
    \label{lemma:para_policy_gradient}
    In MV-TSGs, for a joint policy parameterized by $\theta$, we have
    \begin{equation*}
        \nabla_{\theta} J(\theta) = \mathbb{E}_{s\sim\pi^\theta, \bm{a}\sim \mu^\theta}[\nabla_\theta \log \mu^\theta(\bm{a}|s)Q_f^\theta(s,\bm{a})].
    \end{equation*}
\end{lemma}
The proof is provided in Appendix~\ref{proof:lemma_policy_gradient}. This result is similar to the well-known policy gradient theorem in the average-reward setting \citep{sutton1999policy}, except that the action-value function is computed with respect to the surrogate reward function $f$ instead of 
$r$. Based on Lemma~\ref{lemma:para_policy_gradient}, we further derive the partial derivative formula of the mean-variance performance function. The proof is provided in Appendix~\ref{proof:lemma_perf_derivative}.
\begin{lemma}[Performance partial derivative]
    \label{lemma:para_perf_derivative}
    For any agent $i$ and any directly parameterized policy $\theta_i \in \mathcal{X}_i$, we have
    \begin{equation*}
        \frac{\partial J(\theta)}{\partial \theta_{s,a_i}} = \overline{Q_{f,i}^\theta}(s,a_i)\pi^{\theta}(s),
    \end{equation*}
    where $\overline{Q^\theta_{f,i}}(s,a_i):=\sum\limits_{\bm{a}_{-i}}  \bm\mu^{\theta_{-i}}(\bm{a}_{-i}|s) Q_{f}^{\theta}(s,a_i,\bm{a}_{-i})$.
\end{lemma}
Based on Lemma~\ref{lemma:para_policy_gradient} and Lemma~\ref{lemma:para_perf_derivative}, we now present the independent policy gradient method and analyze its convergence properties.

\subsection{Exact policy gradient ascent}
Because for each agent $i$, the policy parameters should satisfy the constraint of $\sum\limits_{a_i \in \mathcal{A}_i} \theta_{s,a_i}=1, \forall s \in \mathcal{S}$, we propose an independent projected gradient ascent to update policies in MV-TSGs, i.e., MV-IPGA. Specifically, each agent update its policy along the gradient direction and use the operator $\text{Proj}_{\mathcal{X}_i}(\theta_i):=\arg\min\limits_{x \in \mathcal{X}_i}\|x-\theta_i \|$ to project the updated parameters onto the feasible region $\mathcal{X}_i$, 
\begin{equation*}
    \theta_i'=\text{Proj}_{\mathcal{X}_i}(\theta_i+\alpha \nabla_{\theta_i}J(\theta_i, \theta_{-i})), \quad \alpha >0.
\end{equation*}

In MV-TSGs, we have the following lemma. Similar results have also been provided in \cite{leonardos2022global}(Proposition~B.1) and \cite{zhang2024gradient}(Proposition~1).
\begin{lemma}
    \label{lemma:indpg_jotpg}
    Let $\theta=(\theta_1,\dots,\theta_N)$, $\theta' =\theta+\alpha \nabla_{\theta} J(\theta)$, where $\alpha$ is the step size. Then, we have
    \begin{equation*}
        \textnormal{Proj}_{\mathcal{X}}(\theta') = (\textnormal{Proj}_{\mathcal{X}_1}(\theta'_1),\dots,\textnormal{Proj}_{\mathcal{X}_N}(\theta'_N)).
    \end{equation*}
\end{lemma}
Lemma~\ref{lemma:indpg_jotpg} indicates that when all agents adopt the same step size for policy updates, the independent projected gradient ascent is equivalent to performing projected gradient ascent on the joint policy. Based on this observation, we propose the MV-IPGA algorithm, as detailed in Algorithm~\ref{alg:independent_pg}.

\begin{algorithm}[H]
    \caption{Mean-variance independent projected gradient ascent for MV-TSGs}
    \begin{algorithmic}[1]
        \STATE \textbf{Input:} step size $\alpha>0$, number of iterations $K$.
    
        \STATE \textbf{Initialization:} for each agent $i$, randomly initialize the parameters $\theta_i^{(0)}$ of the policy $\mu_i^{(0)}$.
    
        \FOR{$k=0,1,\dots,K-1$}
            \STATE $\theta_i^{(k+1)}=\textnormal{Proj}_{\mathcal{X}_i}(\theta_i^{(k)}+\alpha \nabla_{\theta_i}J(\theta_i^{(k)}, \theta_{-i}^{(k)}))$,\quad $\forall i$.
        \ENDFOR
    \end{algorithmic}
    \label{alg:independent_pg}
\end{algorithm}

However, as demonstrated by \cite{zhong2024heterogeneous}, independent and simultaneous policy updates by all agents may fail to guarantee performance improvement or algorithmic convergence, due to the issue of environmental non-stationarity in multi-agent settings. In light of this, we provide a convergence analysis of MV-IPGA.

\subsection{Global convergence of policy gradient in MV-TSGs}
Based on Lemma~\ref{lemma:indpg_jotpg}, we analyze the convergence of MV-IPGA by following the standard approach for gradient-based methods in nonconvex optimization. The analysis proceeds in two steps: (1) establishing the smoothness properties of the mean-variance performance function, and (2) proving the convergence of MV-IPGA. Without loss of generality, we assume throughout this section that the rewards are normalized such that $r \in [0,1]$.

We first analyze the smoothness properties of the mean-variance performance function, as Lemma~\ref{lemma:smoonth} demonstrates. The proof is provided in Appendix~\ref{proof:lemma_smooth}.
\begin{lemma}\label{lemma:smoonth}
    Denote $A_{\textnormal{max}}:=\max\limits_i |\mathcal{A}_i|$, $S=|\mathcal{S}|$, $\kappa_0$ is the mixing coefficient defined in \cite{cheng2024provable}, $L:= 6\beta A_{\textnormal{max}}(1+\frac{\kappa_0}{2}S)^2 + (1+\beta)\kappa_0A_{\textnormal{max}} \sqrt{S} (\kappa_0 S + 1)$ and $L_J:= N\big(6\beta  A_{\textnormal{max}}(1+\frac{\kappa_0}{2}S)^2 + (1+\beta) A_{\textnormal{max}} (\kappa_0 S+\kappa_0^2 S^{\frac{3}{2}} +\kappa_0 \sqrt{S} + 1)\big)$, we have 
    \begin{itemize}
        \item For any agent $i$ and $\mu_{-i}\in \mathcal{U}_{-i}$, the mean-variance performance function $J^{\theta}$ is $L$-smooth with respect to policy $\theta_i$, i.e., $\left\| \nabla_{\theta_i}J(\theta_i,\theta_{-i}) - \nabla_{\theta_i}J(\theta_i',\theta_{-i})  \right\|_2 \le L \left\| \theta_i - \theta_i' \right\|$, $\forall i$ and $\theta_i,\theta_i' \in \mathcal{X}_i$.
        \item The mean-variance performance function $J^{\theta}$ is $L_J$-smooth with respect to the joint policy $\theta$, i.e., $\left\| \nabla_{\theta}J(\theta) - \nabla_{\theta}J(\theta')  \right\|_2 \le L_J \left\| \theta - \theta' \right\|$.
    \end{itemize}
\end{lemma}
Compared with the smoothness results of the average-reward objective $\eta$ in \cite{cheng2024provable}, it is evident that the incorporation of variance complicates the smoothness of the objective function, which is also influenced by the coefficient $\beta$.

Based on the results of the smoothness properties above, we now analyze the convergence of MV-IPGA. To quantify the convergence behavior of the algorithm, we introduce the function $\text{ST}(\theta)=\max\limits_i \max\limits_{\theta_i'} (\theta_i' - \theta_i)^\top \nabla_{\theta_i}J(\theta)$, which measures the maximal directional improvement over all agents. Since both the long-run average reward $\eta^{\bm\mu}$ and long-run variance $\zeta^{\bm\mu}$ are bounded due to the boundedness of the reward function $r$, we denote the maximum and minimum of the mean-variance performance function by $J_{\max}$ and $J_{\min}$. The convergence result of MV-IPGA is stated in Theorem~\ref{theorem:converge_rate}, and its proof is provided in Appendix~\ref{proof:theorem_converge_rate}.
\begin{theorem}
    \label{theorem:converge_rate}
    If all agents follow  Algorithm~\ref{alg:independent_pg} with step size $\alpha= \frac{1}{L_J}$, then the joint policy asymptotically converges to a first-order stationary point, i.e., $\lim\limits_{k \to \infty} \text{ST}(\theta^{(k)}) \le 0$. Moreover, if all agents independently play gradient ascent for at least $K \ge \frac{4 L_J (J_{\max}-J_{\min})}{\epsilon}$ iterations, then there exists a $k \in \{1,\dots,K \}$ such that $\theta^{(k)}$ is $\epsilon$-stationary, i.e., $\text{ST}(\theta^{(k)}) \le \epsilon$.
\end{theorem}

Theorem~\ref{theorem:converge_rate} shows that MV-IPGA converges within a finite number of iterations when all agents adopt an identical and theoretically valid step size. However, such theoretically guaranteed step sizes are often overly conservative and may be impractical for specific problem instances. In practice, the step size is typically chosen empirically to balance convergence stability and speed.

\section{Independent deep reinforcement learning for MV-TSGs}
\label{sec:MV-IPPO}
In the previous section, we introduced the MV-IPGA algorithm and analyzed its convergence properties. However, applying MV-IPGA to practical energy management in MMSs may encounter two limitations. First, policy gradient methods typically require accurate knowledge of environmental parameters, such as the reward function and state transition dynamics. Second, as the number of microgrids increases, the state space of the system grows exponentially, leading to significant computational and memory challenges for policy optimization based on direct policy parameterization.

To approximately solve the energy management problem in large-scale MMSs under scenarios with unknown model parameters, we further propose an independent DRL approach for MV-TSGs, built upon the independent policy gradient algorithm. In DRL, both the policy and value functions are parameterized via neural networks, which enables the framework to effectively handle problems with high-dimensional state and action spaces. 

In the single-agent setting, PPO has demonstrated remarkable success in practical applications, including OpenAI Five \citep{berner2019dota} and generative models such as ChatGPT. In practical applications, directly applying policy gradient methods may result in excessively large policy updates due to potential parameter estimation errors, thereby destabilizing the learning process. Like other first-order policy gradient algorithms, PPO employs gradient information to guide optimization. The principal difference from methods such as A2C \citep{mnih2016asynchronous}, TD3 \citep{fujimoto2018addressing}, and SAC \citep{haarnoja2018soft} is its conservative update mechanism, which maximizes a clipped surrogate objective to constrain policy updates within a local trust region, thereby enhancing stability. Empirical studies have shown that this approach achieves strong performance across a wide range of tasks, while offering several practical advantages such as ease of implementation, high sample efficiency, and robustness to hyper-parameter settings \citep{schulman2017proximal}.

Following the route of PPO,  we propose a corresponding independent DRL algorithm based on Algorithm~\ref{alg:independent_pg}, referred to as MV-IPPO. Unlike standard PPO, which maximizes the discounted cumulative reward, in MV-IPPO, each agent performs policy optimization based on its surrogate action-value function $\overline{Q_{f,i}^\theta}(s,a_i)$ introduced in Lemma~\ref{lemma:para_perf_derivative} or the advantage function $\overline{A_{f,i}^\theta}(s,a_i):=\sum\limits_{\bm{a}_{-i}}  \bm\mu^{\theta_{-i}}(\bm{a}_{-i}|s) A_{f}^{\theta}(s,a_i,\bm{a}_{-i})$. At the $k$th iteration, the policy of each agent and the common value function are parameterized by neural networks with parameters $\theta_i^{(k)}$ (actor network) and $\phi^{(k)}$ (critic network), respectively. Then, each agent optimizes its policy by maximizing the following surrogate objective function 
\begin{equation*}
    \mathcal{L}^{\bm\mu}_{i}(\theta_i):=\mathbb{E}_{s\sim \pi^{\bm\mu},a_i\sim \mu_i}\left[\min\left(\omega_i(\theta_i)\hat{A}_{f,i}^{\theta^{(k)}}(s,a_i),\mathrm{clip}(\omega_i(\theta_i),1-\varepsilon,1+\varepsilon)\hat{A}_{f,i}^{\theta^{(k)}}(s,a_i)\right)\right],
\end{equation*}
where $\omega_i(\theta_i)=\frac{\mu^{\theta_i}(a_{i}|s)}{\mu^{\theta_{i}^{(k)}}(a_{i}|s)}$, and $\hat{A}_{f,i}^{\theta^{(k)}}$ is the estimation of $\overline{A_{f,i}^{\theta^{(k)}}}$.

For each agent, the advantage function $\hat{A}_{f,i}$ can be estimated using the generalized advantage estimation method \citep{schulman2015high}. Specifically, we have
\begin{equation*}
        \hat{A}_{f,i}^{{\theta^{(k)}}}(s_n,a_{i,n})=\sum\limits_{t=n}^{T-1}\lambda_{t-n}(\hat{f}^{\theta^{(k)}}(s_t,a_{i,t}) - \hat{J}^{{{\theta^{(k)}}}}+V_f^{\phi^{(k)}}(s_t) - V_f^{\phi^{(k)}}(s_{t+1})),
    \label{eq:compute_advantage}
\end{equation*}
where $\lambda$ is the hyper-parameter to trade-off bias and variance, and $T$ denotes the trajectory length. 

Since we consider the long-run average performance in this study, we adopt the average value constraint (AVC) proposed by \cite{maaverage} to assist in estimating the target value function $\hat{V}_f^{\phi^{(k)}}$ and stabilizing the value learning. The value function network is updated using the loss function
\begin{equation}
     \frac{1}{T}\sum\limits_{t=0}^{T-1} \big( V_f^{\phi}(s_t) - \hat{V}_f^{\phi^{(k)}}(s_t)  \big)^2. \notag
\end{equation}
The detailed pseudo-code of MV-IPPO is presented in Algorithm~\ref{alg:independent_ppo}, where $(\cdot)_{t,m}$  denotes the corresponding variable value at time step $t$ in the $m$-th trajectory.

\begin{algorithm}[H]
    \caption{Independent proximal policy optimization for MV-TSGs}
    \setstretch{0.9}
    \begin{algorithmic}[1]
        \STATE \textbf{Input:} threshold hyper-parameter $\epsilon>0$, step size $\alpha$, episode length $T$.
    
        \STATE \textbf{Initialization:} for agent $i$, randomly initialize the parameters $\theta_i^{(0)}$ of the policy $\mu_i^{(0)}$ and the parameters $\phi^{(0)}$ of the value function, replay buffer $\mathcal{M}$. $M$ is the total number of trajectories collected in each iteration. Set $\hat{\eta}=0, \hat{\zeta}=0, \hat{J}=0$.
    
        \FOR{$k=0,1,\dots$}
            \STATE Collect a set of trajectories by running policy $\mu^{\theta_i}$ in the environment.
            \STATE Push transitions $\{(s_{t,m},a_{i,t,m} ,s_{t+1,m},r_{t,m}), t \in \{0,\dots,T-1\},m\in\{1,\dots,M\}\}$ into $\mathcal{M}$.
            \STATE Update $\hat{\eta} \gets (1-\alpha)\hat{\eta}+\alpha\frac{1}{MT} \sum\limits_{m=1}^{M} \sum\limits_{t=0}^{T-1}r_{t,m} $.

            \STATE Update $\hat{\zeta} \gets (1-\alpha)\hat{\zeta}+\alpha\frac{1}{MT} \sum\limits_{m=1}^M \sum\limits_{t=0}^{T-1}(r_{t,m}-\hat{\eta})^2 $. 

            \STATE Compute the average mean-variance performance function $\hat{J}$.   
        
            \STATE Compute the average $\hat{f}^{\theta^{(k)}}(s_t, a_{i,t})$ and $\hat{A}_f^{{\theta^{(k)}}}(s_t, a_{i,t})$ over all time steps and trajectories.
        
            \STATE Compute the average $\hat{V}_f^{\phi^{(k)}}(s_t)$ over all time steps and trajectories using AVC.

            \STATE Update the policy by maximizing the objective:
            \begin{align*}
            \theta_i^{k+1}=\arg\max\limits_{\theta_i} \frac{1}{MT}  \sum_{m=1}^M\sum_{t=0}^{T-1}\operatorname*{min}\Big(& \frac{\mu^{\theta_i}(a_{i,t,m}|s_{t,m})}{\mu^{\theta_{i}^{(k)}}(a_{i,t,m}|s_{t,m})}\hat{A}_{f,i}^{\theta^{{k}}}(s_{t,m},a_{i,t,m}), \\
                &\text{clip}\Big(\frac{\mu^{\theta_i}(a_{i,t,m}|s_{t,m})}{\mu^{\theta_{i}^{(k)}}(a_{i,t,m}|s_{t,m})}, 1\pm \epsilon\Big)\hat{A}_{f,i}^{\theta^{{k}}}(s_{t,m},a_{i,t,m})\Big),
            \end{align*}
            using stochastic gradient ascent with Adam.
            \STATE Fit value function by regression on mean-squared error:
            \begin{equation*}
            \phi_{k+1} = \mathop{\arg\min}_{\phi} \frac{1}{MT}\sum\limits_{m=1}^M\sum\limits_{t=0}^{T-1} \big( V_f^{\phi}(s_{t,m}) - \hat{V}_f^{\phi^{(k)}}(s_{t,m})  \big)^2,
            \end{equation*} 
            using stochastic gradient descent with Adam.
        \ENDFOR
    \end{algorithmic}
    \label{alg:independent_ppo}
\end{algorithm}

\section{Numerical experiments}{\label{sec:experiments}}
This section first introduces the experimental settings in MMSs. We then evaluate the effectiveness of MV-IPGA under varying values of $\beta$ in a simplified scenario. Finally, MV-IPPO is applied to address the energy management problem in a larger-scale setting with unknown model parameters.  

\subsection{Experimental settings}
\noindent We investigate an MMS under the distributed EMS scheme, specifying the configurations of renewable generators, demand loads, and ESSs based on typical equipment and power demand profiles \citep{yang2020fluctuation}. For simplicity, we assume that the equipment in different microgrids is identical and operates independently. All state and action variables are discretized using uniform quantization.

(1) Renewable energy generator. A wind turbine serves as a representative renewable energy generator in this study. The wind power $G(t)$ are calculated by 
\begin{equation*}
    G(t)=\begin{cases}
W^{\textrm{cap}}, & V^{\textrm{rated}}\leq v_{t}<V^{\textrm{cutout}}, \\
W^{\textrm{cap}}\left(\frac{v_{t}}{V^{\textrm{rated}}}\right)^{3}, & V^{\textrm{cutin}}\leq v_{t}<V^{\textrm{rated}}, \\
0, & \text{others},
\end{cases}
\end{equation*}
where $v_t$ represents the wind speed at time step $t$, $V^{\text{rated}}$, $V^{\textrm{cutin}}$ and $V^{\textrm{cutout}}$ indicate the rated, cut-in and cut-out wind speed, respectively. $W^{\textrm{cap}}$ denotes the rated wind power. The parameter settings of the wind turbine are presented in Table~\ref{tab:config_windturbine}. 

\begin{table}[!htbp]
    \begin{center}
    \caption{Wind turbine configurations.}
        \begin{tabular}{ccc}
        \hline
            Symbol  & Description & Setting \\
        \hline
            $V^{\textrm{cutin}}$  & Cut-in wind speed & 4 m/s \\
        \hline
             $V^{\textrm{cutout}}$ & Cut-out wind speed & 25 m/s \\

        \hline
            $V^{\text{rated}}$  & Rated wind speed & 15 m/s \\
        \hline
            $W^{\textrm{cap}}$  & Rated wind power & 3 MW \\
        \hline
        \end{tabular}
    \label{tab:config_windturbine}
    \end{center}
\end{table}

\begin{table}[htbp]
    \begin{center}
    \caption{States of wind power.}
        \begin{tabular}{ccccccc}
        \hline
            State  & 1 & 2 & 3 & 4 & 5 & 6  \\
        \hline
            Wind power/MW & 0 & 0.6 & 1.2 & 1.8 & 2.4 & 3.0 \\
        \hline
        \end{tabular}
    \label{tab:state_G}
    \end{center}
\end{table}

\begin{equation}\label{eq:P_g}
\bm{P}_W= \left(
\begin{matrix}
0.94 & 0.05 & 0.01 & 0.00 & 0.00 & 0.00\\
0.40 & 0.44 & 0.10 & 0.03 & 0.01 & 0.02\\
0.16 & 0.37 & 0.26 & 0.11 & 0.05 & 0.05\\
0.08 & 0.24 & 0.25 & 0.19 & 0.10 & 0.14\\
0.08 & 0.14 & 0.18 & 0.19 & 0.14 & 0.27\\
0.04 & 0.07 & 0.08 & 0.10 & 0.10 & 0.61
\end{matrix}
\right).
\end{equation}

The output power of the wind turbine is discretized into six states, as illustrated in Table~\ref{tab:state_G}. The wind power and its state transition are determined by the random wind speed. The wind speed data we used were collected by the Measurement and Instrumentation Data Center at the National Renewable Energy Laboratory over the period from 2015 to 2024 \citep{NREL2025}. Accordingly, the probability transition matrix $\bm{P}_W$ in (\ref{eq:P_g}) is constructed. The $(k,l)^{\text{th}}$ element $p_{k,l}$ in $\bm{P}_W$ indicates the probability that the wind power from state $k$ to state $l$, and is calculated by $P_{k,l}=\frac{q_{k,l}}{q_k}$, where $q_{k,l}$ represents the observed transitions from state $k$ to state $l$, and $q_k$ represents the total occurrence of state $k$.

(2) Demand load. The demand load data is derived from the 2023 data in a publicly available database \citep{ieso2023}. The database is maintained by the independent electricity system operator (IESO), which is a non-profit corporate entity established in 1998 by the Electricity Act of Ontario. The dataset employed records the hourly demand load in Ontario across the entire year. With this demand load dataset, we divide the demand load power into six states and construct the probability transition matrix $\bm{P}_D$, as illustrated in Table~\ref{tab:state_D} and (\ref{eq:P_d}), in the same way as in the wind power.

\begin{table}[htbp]
    \begin{center}
    \caption{States of demand loads.}
        \begin{tabular}{ccccccc}
        \hline
            State  & 1 & 2 & 3 & 4 & 5 & 6  \\
        \hline
            Demand load power/MW & 0.6 & 1.2 & 1.8 & 2.4 & 3.0 & 3.6 \\
        \hline
        \end{tabular}
    \label{tab:state_D}
    \end{center}
\end{table}

\begin{equation}\label{eq:P_d}
\bm{P}_D= \left(
\begin{matrix}
0.75 & 0.25 & 0.00 & 0.00 & 0.00 & 0.00\\
0.03 & 0.83 & 0.14 & 0.00 & 0.00 & 0.00\\
0.00 & 0.11 & 0.82 & 0.07 & 0.00 & 0.00\\
0.00 & 0.00 & 0.14 & 0.84 & 0.02 & 0.00\\
0.00 & 0.00 & 0.00 & 0.19 & 0.79 & 0.02\\
0.00 & 0.00 & 0.00 & 0.00 & 0.27 & 0.73
\end{matrix}
\right).
\end{equation}

(3) Energy storage system. The parameters of the ESS are presented in Table~\ref{tab:config_storage}. The energy level is divided into six states, as illustrated in Table~\ref{tab:state_storage}. The action space for discharging power is also discretized properly and illustrated in Table~\ref{tab:action_storage}.

\begin{table}[htbp]
    \begin{center}
    \caption{Configurations of energy storage systems.}
        \begin{tabular}{ccc}
        \hline
            Symbol  & Description & Setting \\
        \hline
            $C^{\textrm{ch}}$  & Maximum charging power & 1.2 MW \\
        \hline
             $C^{\textrm{dis}}$ & Minimum discharging power & 1.2 MW \\
        \hline
            $B^{\text{cap}}$  & Rated capacity & 4.0 MWh \\
        \hline
            $B^{\textrm{min}}$  & Energy level lower bound threshold & 0.6 MWh \\
        \hline
            $B^{\textrm{max}}$  & Energy level upper bound threshold & 3.6 MWh \\
        \hline
            $\nu$  & Charging/discharging efficiency coefficient & 0.95 \\
        \hline
        \end{tabular}
    \label{tab:config_storage}
    \end{center}
\end{table}

\begin{table}[htbp]
    \begin{center}
    \caption{States of storage energy levels.}
        \begin{tabular}{ccccccc}
        \hline
            State  & 1 & 2 & 3 & 4 & 5 & 6  \\
        \hline
            Energy level/MWh & 0.6 & 1.2 & 1.8 & 2.4 & 3.0 & 3.6\\
        \hline
        \end{tabular}
    \label{tab:state_storage}
    \end{center}
\end{table}

\begin{table}[htbp]
    \begin{center}
    \caption{Scheduling actions of energy storage systems.}
        \begin{tabular}{cccccc}
        \hline
            Action ($b_{i,t}$)  & 1 & 2 & 3 & 4 & 5   \\
        \hline
            Discharging power/MW & -1.2 & 0.6 & 0.0 & 0.6 & 1.2 \\
        \hline
        \end{tabular}
    \label{tab:action_storage}
    \end{center}
\end{table}

We conduct a series of experiments to validate the effectiveness of the MV-IPGA and MV-IPPO algorithms in Section~\ref{sec:exp_model_based} and Section~\ref{sec:exp_model_free}, respectively. All experiments are conducted on a machine equipped with one AMD 3995WX CPU, 384GB of memory, and one Nvidia GeForce GTX4090 GPU. We note that the corresponding algorithms are evaluated under different values of $\beta$, and compared against a baseline policy without energy management, due to the lack of existing methods that can address the economic and reliable MMS energy management under the distributed EMS scheme.

\subsection{Policy gradient in a small-scale  case}\label{sec:exp_model_based}
In this section, we use the MV-IPGA method to deal with the economic and reliable energy management problem when the environmental parameters are known exactly. Due to the state space increasing exponentially with the number of microgrids, we focus on a small-scale MMS consisting of two microgrids. Each microgrid is equipped with an ESS, while only Microgrid 1 is configured with a renewable energy generator and demand loads. The step size $\alpha$ in the policy gradient method is set to 0.5.

\begin{figure}[!htbp] 
    \centering \includegraphics[width=16cm]{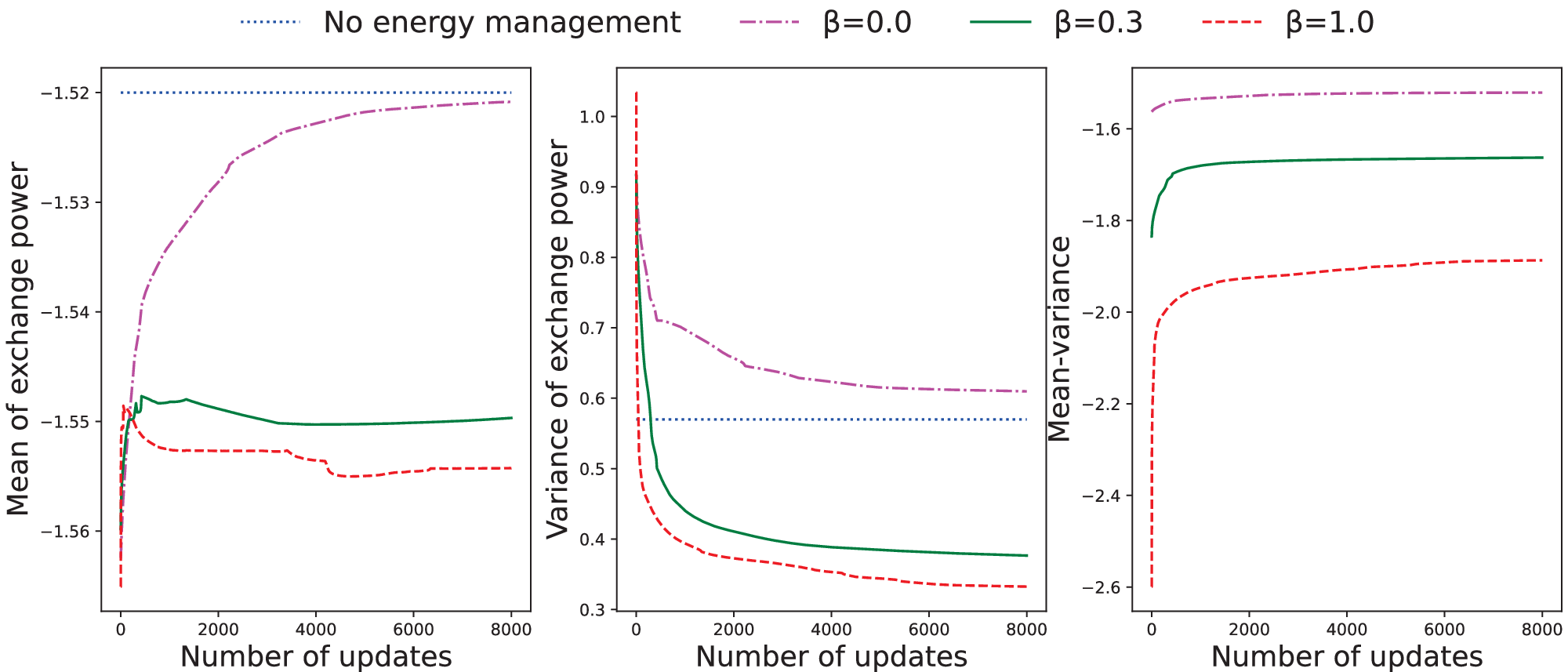} \caption{Convergence procedure of MV-PG with different coefficients.} 
    \label{figure:convergence_pg} 
\end{figure}

Figure~\ref{figure:convergence_pg} illustrates the convergence process of the mean, variance, and mean-variance of exchange power under MV-IPGA. In the first two subplots, the blue dashed lines indicate the corresponding values under a baseline policy without energy management, with a mean of -1.52 and a variance of 0.57. Since energy management involves ESS scheduling (with efficiency coefficient $\nu<1$) or power curtailment, the blue dashed line in the first subplot represents the theoretical upper bound of the mean exchange power. To evaluate the effectiveness of MV-IPGA, we conduct experiments under different values of $\beta$: 0.0, 0.3, and 1.0. As shown in the figure, when $\beta=0.0$, the mean exchange power gradually converges to -1.52. In this case, the algorithm does not account for power fluctuation minimization, and
the variance remains above the baseline level. We note that the variance decreases because the mean optimization avoids unnecessary energy management actions.

When $\beta=0.3$, the power fluctuation is taken into account, and the converged variance of exchange power reduces significantly to 0.38, though the converged mean value incurs a slight decrease due to energy losses associated with storage scheduling and curtailment.  When $\beta=1.0$, more considerations are placed on reducing power fluctuations, further adjusting both the variance and mean. 

The convergence curves of the mean-variance objective function in the third subplot clearly show that the proposed objective is progressively optimized as the number of algorithm iterations increases, demonstrating the effectiveness of MV-IPGA.

\begin{figure}[!htbp] 
    \centering \includegraphics[width=16cm]{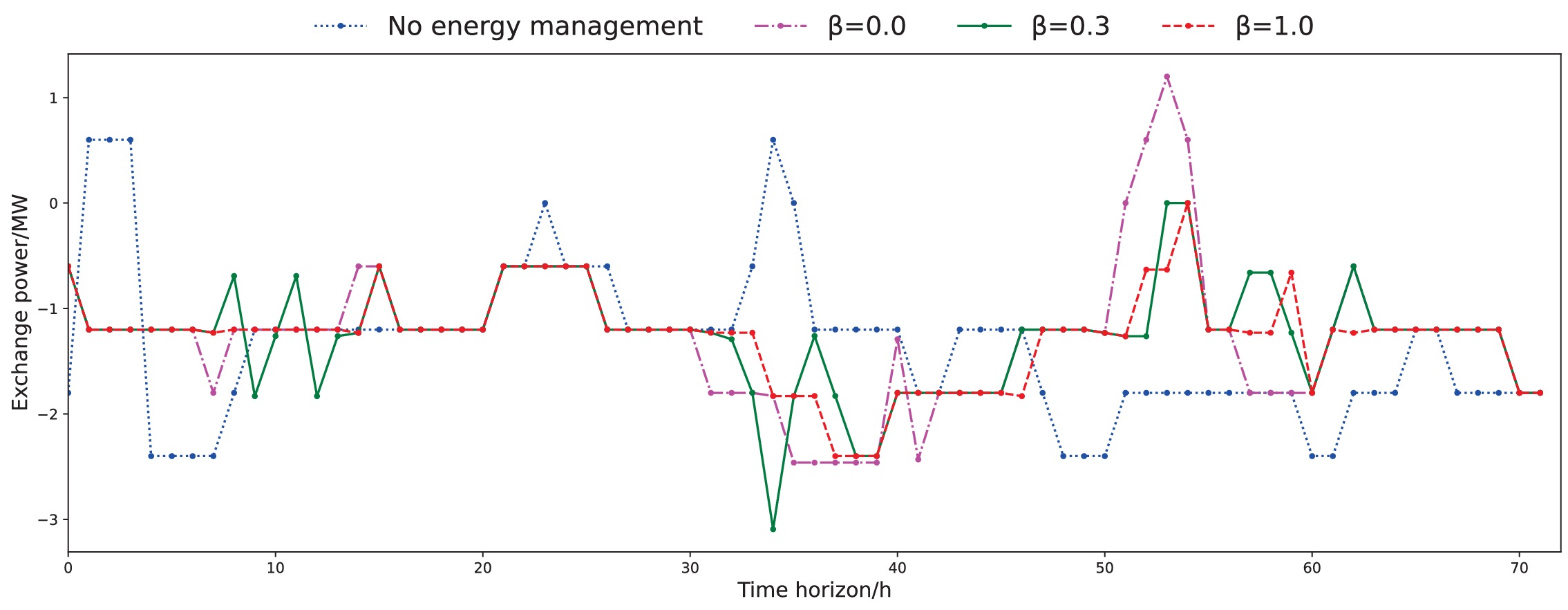} \caption{An episode of exchange power over 72 time steps under the MV-IPGA policy.} 
    \label{figure:episode_pg} 
\end{figure}

To provide a more intuitive demonstration of the superiority of the energy management policy learned by the algorithm, a sample trajectory of exchange power over 72 time steps is plotted in Figure~\ref{figure:episode_pg}. It is illustrated that the exchange power exhibits significant fluctuations when no energy management is applied (blue curve) or when power fluctuation is not considered in the optimization (purple curve). In contrast, when $\beta$ is set to 0.3 and 1.0, the fluctuations in exchange power are effectively mitigated to varying degrees. 

\begin{figure}[!htbp] 
    \centering \includegraphics[width=16cm]{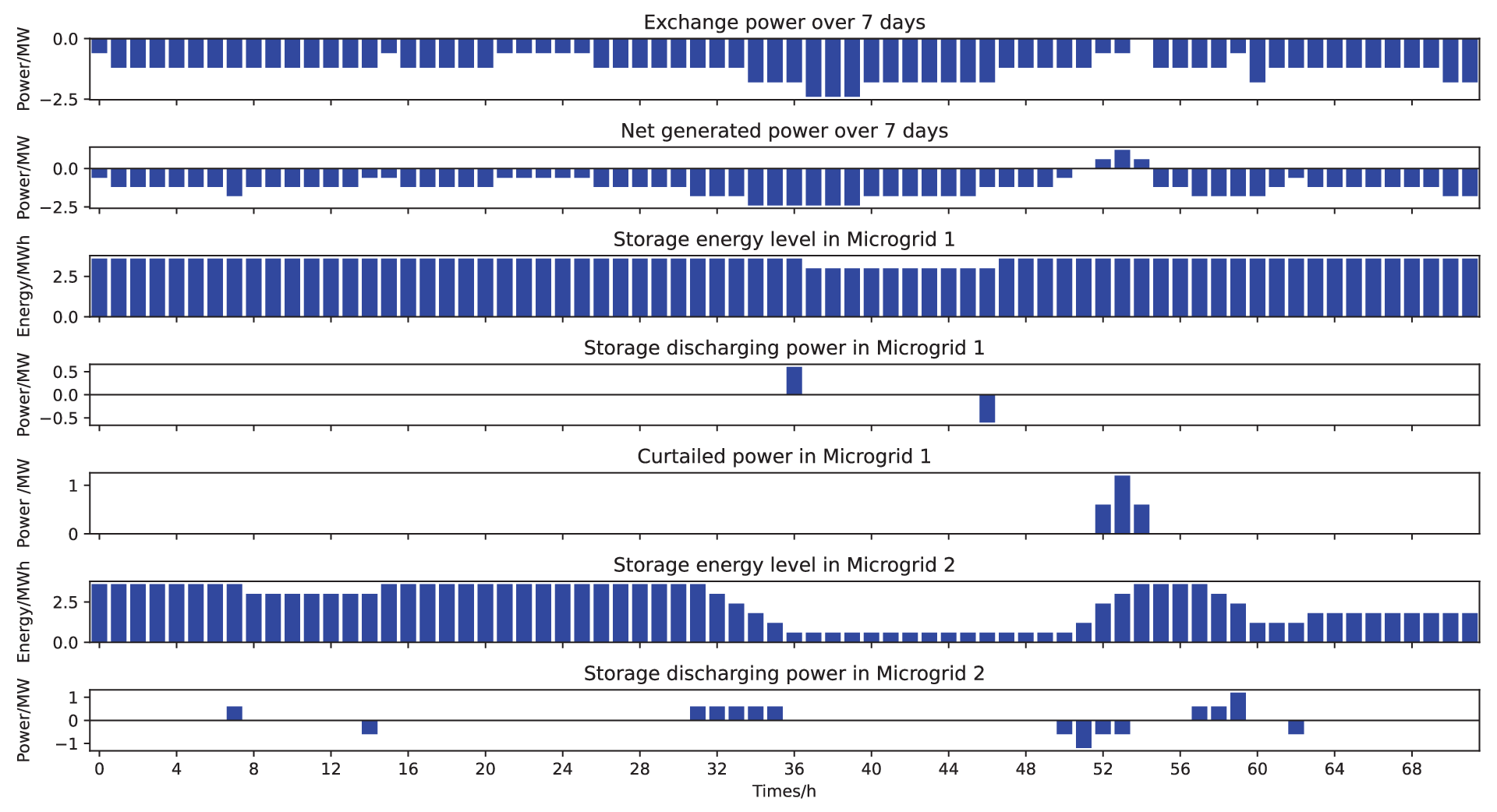} \caption{Trajectory details of states and actions when $\beta=1.0$.} 
    \label{figure:episode_details} 
\end{figure}

Furthermore, Figure~\ref{figure:episode_details} presents the state and action details of a representative trajectory under the setting $\beta=1.0$. The net generated power is defined as the difference between the generated wind power and the demand load, i.e., $G_{1,t}-D_{1,t}$. It can be observed that the net generated power remains negative for most time steps, indicating that the demand load generally exceeds the available wind power in MMS operations. Microgrids only take action in states where significant fluctuations are likely to occur, as the discharging and charging of storage, as well as power curtailment, can result in energy losses.

Specifically, when the demand significantly surpasses the generated power, the microgrids compensate for power by discharging ESSs, such as during time steps 31 to 36. Conversely, microgrids tend to charge their ESSs when the net power is only slightly negative or turns positive. Notably, when the net power becomes positive, Microgrid 1 implements power curtailment to prevent reverse power flow and to mitigate power fluctuations. Such behavior can be observed between time steps 50 and 54.

In summary, the results presented in the above figures jointly demonstrate the effectiveness of the proposed MV-IPGA in addressing the trade-off between economic performance and reliability in energy management for MMSs.

\subsection{Reinforcement learning in a large-scale case}\label{sec:exp_model_free}
In this section, we use MV-IPPO to tackle the energy management problem when environmental parameters are unknown. The MMS considered consists of three microgrids, all are equipped with renewable energy generators, demand loads, and ESSs. 

A simulated environment is constructed to train the energy management policy of microgrids. At each time step, the microgrid takes an action $a_{i,t}$ to interact with the environment and get a reward $r_{i,t}$ feedback. The system state $s_t$ is transitioned to $s_{t+1}$ according to the transition matrix $P_W$, $P_D$, and the joint action $\bm{a}$. The microgrids iteratively update their energy management policies based on the interaction data with the environment. The primary hyper-parameters of MV-IPPO are presented in Table~\ref{tab:hyper-para}. Both the actor and critic networks comprise two hidden layers, each with 64 units and rectified linear unit (ReLU) activation functions. The network architecture and most hyperparameter settings follow those of the state-of-the-art algorithm in \cite{yu2022surprising}.

 \begin{table}[htbp]
        \begin{center}
        \caption{Hyper-parameter settings in MV-IPPO}
            \begin{tabular}{ccc}
            \hline
                Hyper-parameter  & Value\\
            \hline
                Number of training steps  & 20 million  
                  \\  
                Number of environments collecting data in parallel  & 20 \\
                Length of time horizon $T$  &   2000\\
                Decay-rate parameter $\lambda$ for eligibility traces & 0.95 \\ 
                Number of mini-bath  & 40 \\
                Clipping coefficient $\epsilon$ & 0.2 \\ 
                Training epochs &   5\\
                Optimizer for gradient descent/ascent & Adam  \\ 
                Learning rate for optimizer  &0.0005    \\
                Average value constraint coefficient in AVC & 0.01 \\
            \hline
            \end{tabular}
        \label{tab:hyper-para}
        \end{center}
    \end{table}

\begin{figure}[!htbp] 
    \centering \includegraphics[width=16cm]{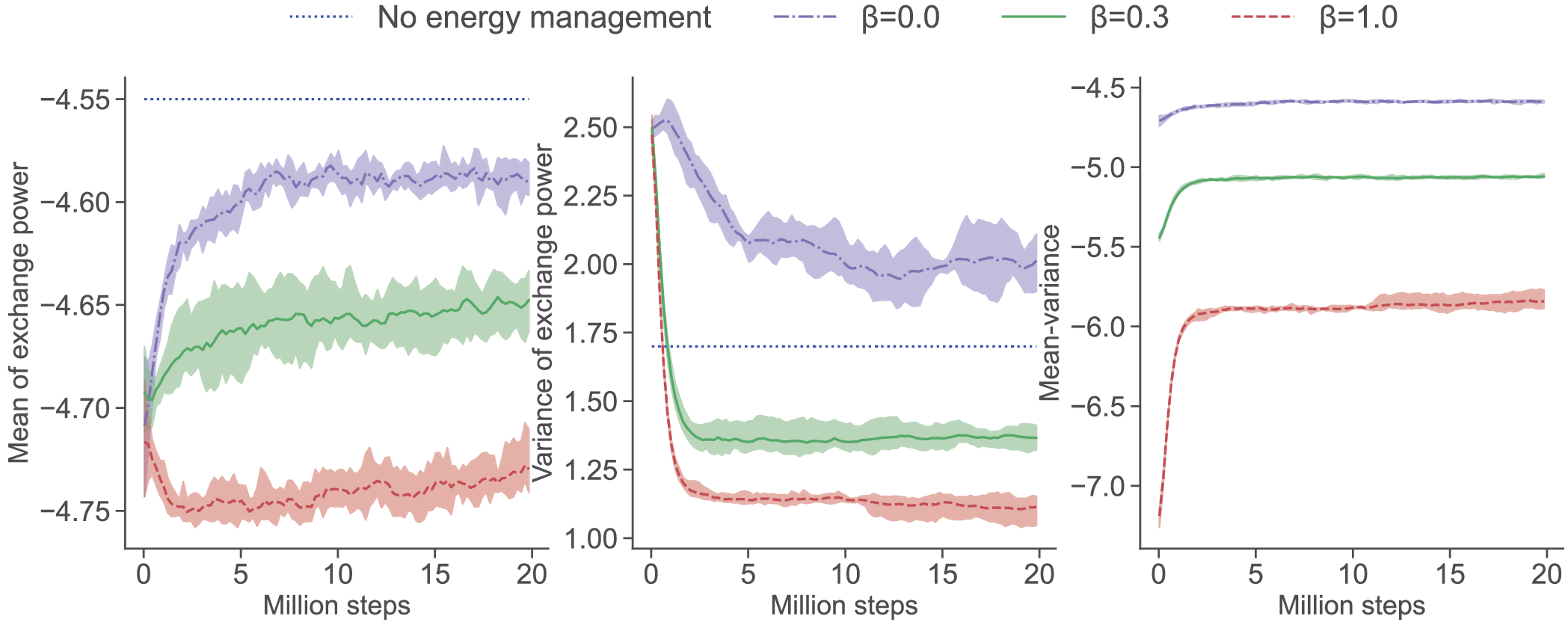} \caption{\textbf{Training curves of MV-IPPO with different coefficients.} Each training curve in the figure is averaged over six random seeds and shaded by standard deviations. } 
    \label{figure:convergence_ppo} 
\end{figure}

The performance of MV-IPPO is evaluated under different values of $\beta$, with the corresponding training curves shown in Figure~\ref{figure:convergence_ppo}. The blue curves represent the long-term mean and variance of exchange power in the absence of energy management, which are $-4.55$ and 1.70, respectively. As illustrated in the figure, when $\beta=0.0$, MV-IPPO focuses solely on optimizing the mean exchange power without accounting for power fluctuations. In this case, the average converged mean across multiple random seeds reaches $-4.59$, which is very close to the optimal value of $-4.55$, while the resulting variance exceeds 1.70. 

\vspace{-1em}
\begin{figure}[!htbp] 
    \centering \includegraphics[width=16cm]{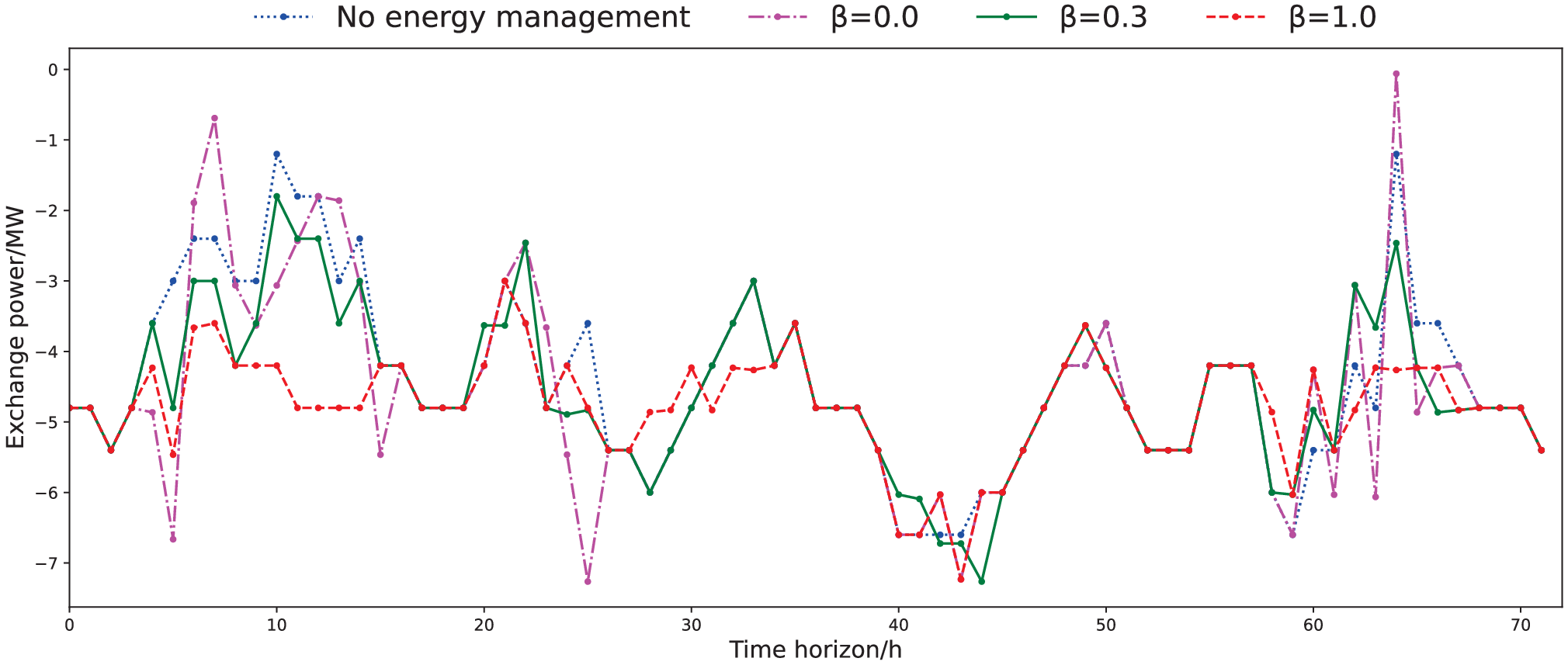} \caption{An episode of exchange power over 72 time steps under the MV-IPPO policy.} 
    \label{figure:episode_ppo} 
\end{figure}

When $\beta=0.3$, the variance of exchange power is explicitly considered. As shown in Figure~\ref{figure:convergence_ppo}, the variance is significantly reduced upon convergence, indicating effective coordination between energy storage and renewable generation resources. However, due to energy losses caused by storage charging/discharging and power curtailment, the mean value also decreases. As $\beta$ increases further to 1.0, both the converged mean and variance are further regulated. The convergence curves of the mean-variance objective confirm that the objective function is effectively optimized as training progresses.

Similarly, we select the best-performing policies across different random seeds for evaluation and generate a representative sample trajectory over 72 time steps, as depicted in Figure~\ref{figure:episode_ppo}. It can be intuitively observed from the figure that the fluctuations in exchange power are effectively mitigated as the value of $\beta$ increases.

The results presented in Figure~\ref{figure:convergence_ppo} and Figure~\ref{figure:episode_ppo} demonstrate that MV-IPPO effectively solves large-scale MV-TSG problems in a data-driven manner. The derived policies facilitate the joint optimization of economic performance and reliability in MMSs.

\section{Conclusion and discussion}{\label{sec:conclusion}}
\noindent In this paper, we investigate an economic and reliable energy management problem for MMSs under the distributed EMS scheme. We introduce the mean and variance of exchange power between the MMS and the main grid as indicators of economic performance and reliability, respectively. The problem is formulated as an MV-TSG. Given the absence of centralized coordination in the distributed EMS scheme, we propose an independent policy gradient method, termed MV-IPGA, to solve the MV-TSG when model parameters are fully known. A rigorous convergence analysis of MV-IPGA is also provided. Furthermore, we develop a DRL algorithm, called MV-IPPO, which extends MV-IPGA to scenarios with unknown model parameters. MV-IPPO enables the approximate solution of large-scale MV-TSGs in a data-driven manner. Experimental results demonstrate that the proposed algorithms effectively solve MV-TSGs and achieve a desirable trade-off between economic performance and reliability in the long-term operation of MMSs.

This study develops a distributed optimization framework for mean–variance objectives. Extending it to alternative risk measures, such as value-at-risk (VaR) and CVaR, is a promising direction for future research. However, these risk metrics often lack analytically tractable formulations for performance difference or gradient computation, which are essential to the proposed algorithm. In practice, challenges such as communication delays and missing data may be mitigated by leveraging recurrent neural networks or attention mechanisms to capture temporal dependencies. Moreover, addressing the sim-to-real gap through robustness- and generalization-oriented techniques is crucial to enhance the real-world applicability of DRL policies.

\section*{Funding Declaration}
This work was supported in part by the National Key Research and Development Program of China (2022YFA1004600). %, in part by the National Natural Science Foundation of China (72342006, 72371253), and in part by Guangdong Basic and Applied Basic Research Foundation (2023A1515012492).

\bibliographystyle{apalike2} 
\bibliography{ref}

\newpage
\section*{Appendix}

\appendix

\section{Proofs}

\subsection{Proof of Lemma~\ref{lemma:para_policy_gradient}}
\label{proof:lemma_policy_gradient}

\proof{}
    Consider two joint policies, $\bm\mu'$ and $\bm\mu$, parameterized by $\theta'$ and $\theta$, respectively. According to Lemma~\ref{lemma:para_perf_diff}, we have
    \begin{align*}
        J(\bm{\mu}') - J(\bm{\mu}) 
        = \mathbb{E}_{s \sim \pi^{\bm{\mu}'}, \bm{a} \sim \bm{\mu}'}[A_{f}^{\bm{\mu}}(s,\bm{a})] + \beta(\eta^{\bm\mu'} - \eta^{\bm\mu} )^2 .
    \end{align*}
    
    We denote $h_1(\bm\mu',\bm\mu) = \mathbb{E}_{s \sim \pi^{\bm{\mu}'}, \bm{a} \sim \bm{\mu}'}[A_{f}^{\bm{\mu}}(s,\bm{a})]$ and $h_2(\bm\mu',\bm\mu) = (\eta^{\bm\mu'} - \eta^{\bm\mu} )^2$.
    Considering the policy parameterization, 
    we denote $\Delta \theta = \theta'-\theta$ and 
    let $\theta' \to \theta$. For $h_1(\theta',\theta)$ we have
    \begin{align*}
        \nabla_{\theta} h_1 &= \lim\limits_{\Delta \theta \to 0} \frac{1}{\Delta \theta} \sum\limits_s \pi^{\theta'}(s) \sum\limits_{\bm{a}} \left[ \bm\mu^{\theta'}(\bm{a}|s) A_f^{\theta}(s,\bm{a}) \right] \notag\\
        & \overset{\text{(i)}}{=} \lim\limits_{\Delta \theta \to 0} \sum\limits_s \pi^{\theta'}(s) \sum\limits_{\bm{a}} \frac{\bm\mu^{\theta'}(\bm{a}|s) - \bm\mu^{\theta}(\bm{a}|s)}{\Delta \theta} A_f^{\theta}(s,\bm{a}) \notag\\
        & = \sum\limits_s \pi^{\theta}(s) \sum\limits_{\bm{a}} \nabla_{\theta} \bm\mu^{\theta}(\bm{a}|s) A_f^{\theta}(s,\bm{a}) \notag\\
        & \overset{\text{(ii)}}{=} \mathbb{E}_{s \sim \pi, \bm{a} \sim \bm\mu^{\theta}} \left[  \nabla_{\theta} \log\bm\mu^{\theta}(\bm{a}|s) A_f^{\theta}(s,\bm{a}) \right] \notag\\
        & = \mathbb{E}_{s \sim \pi, \bm{a} \sim \bm\mu^{\theta}} \left[  \nabla_{\theta} \log\bm\mu^{\theta}(\bm{a}|s) Q_f^{\theta}(s,\bm{a}) \right],
    \end{align*}
    where the Equality $\text{(i)}$ holds is due to  $\mathbb{E}_{\bm{a}\sim \bm\mu} [A_f^{\bm\mu}(s,\bm{a})]=0$, Equality $\text{(ii)}$ holds is due to $\nabla_{\theta}\log\bm\mu^{\theta}(\bm{a}|s) = \frac{\nabla_{\theta} \bm\mu^{\theta}(\bm{a}|s)}{\bm\mu^{\theta}(\bm{a}|s)}$. For $h_2(\theta',\theta)$ we have
    \begin{align*}
        \nabla_{\theta} h_2 
        & = \lim\limits_{\Delta \theta \to 0} \frac{(\eta^{\theta'} - \eta^{\theta})^2}{\Delta \theta} \notag\\
        & =  \lim\limits_{\Delta \theta \to 0} 2(\eta^{\theta'}-\eta^{\theta}) \nabla_{\theta}\eta^{\theta} \notag\\
        & = 0.
    \end{align*}
    Then, we arrive at
    \begin{align*}
        \nabla_{\theta} J(\theta) 
        & = \nabla_\theta h_1 - \beta \nabla_\theta h_2 \notag\\
        & = \mathbb{E}_{s \sim \pi^{\theta}, \bm{a} \sim \bm\mu^{\theta}} \left[  \nabla_{\theta} \log\bm\mu^{\theta}(\bm{a}|s) Q_f^{\theta}(s,\bm{a}) \right],
    \end{align*}
    then the proof is finished.
\endproof

\subsection{Proof of Lemma~\ref{lemma:para_perf_derivative}}
\label{proof:lemma_perf_derivative}
\proof{}
According to the policy gradient Lemma~\ref{lemma:para_policy_gradient}, we have
    \begin{equation*}
        \frac{\partial J(\theta)}{\partial \theta_{s,a_i}} = \sum\limits_{\bm{a}} \pi^\theta (s) \bm\mu^\theta (\bm{a}|s) \frac{\partial \log \bm\mu^\theta (\bm{a}|s)}{\partial \theta_{s,a_i}} Q_f^\theta(s,\bm{a}).
    \end{equation*}
    For the direct policy parameterization, with $\mu^{\theta}(\bm{a}|s)=\prod_{i}^N \mu^{\theta_i}(a_i|s)$, we have
    \begin{align*}
        \frac{\partial \log \bm\mu^\theta (\bm{a}|s)}{\partial \theta_{s,a_i}} = \frac{\partial \log \mu^{\theta_i} (a_i|s)}{\partial \theta_{s,a_i}} 
        = \frac{1}{\mu^{\theta_i}(a_i|s)}.
    \end{align*}
    Then we arrive at
    \begin{align*}
        \frac{\partial J(\theta)}{\partial \theta_{s,a_i}} 
        &= \sum\limits_{\bm{a}_{-i}} \pi^{\theta} (s) 
        \mu^{\theta_{-i}}(\bm{a}_{-i}|s) 
        \mu^{\theta_{i}}(a_{i}|s) \frac{1}{\mu^{\theta_i}(a_i|s)} Q_f^{\theta}(s,\bm{a}) \notag\\
        & = \sum\limits_{\bm{a}_{-i}} \pi^{\theta} (s) \mu^{\theta_{-i}}(\bm{a}_{-i}|s) Q_f^{\theta}(s,\bm{a}) \notag\\
        & = \pi^{\theta}(s) \overline{Q_{f,i}^{\theta}}(s,a_i),
    \end{align*}
    where $\overline{Q^\theta_{f,i}}(s,a_i):=\sum\limits_{\bm{a}_{-i}}  \bm\mu^{\theta_{-i}}(\bm{a}_{-i}|s) Q_{f}^{\theta}(s,a_i,\bm{a}_{-i})$.
    The proof is finished.
\endproof

\subsection{Proof of Lemma~\ref{lemma:smoonth}}
\label{proof:lemma_smooth}
\proof{}
First, we provide the following definitions.
        \begin{align*}
            \theta_{i,\epsilon}(a_i|s)&=\theta_{s,a_i}+\epsilon u_{s,a_i},\\
            \theta_{j,\tau}(a_j|s)&=\theta_{s,a_j}+\tau u_{s,a_j} ,\\
            \theta_{\epsilon}(\bm{a}|s)&= \theta_{i,\epsilon}(a_i|s)\theta_{-i}(a_{-i}|s) ,\\
            \theta_{\tau}(\bm{a}|s)&= \theta_{j,\tau}(a_j|s)\theta_{-j}(a_{-j}|s) ,\\
            J_{\epsilon}&=J(\theta_{i,\epsilon},\theta_{-i}) ,\\
            J_{\epsilon,\tau}&=J(\theta_{i,\epsilon},\theta_{j,\tau},\theta_{-ij}), \\
            \overline{r_i^\theta}(s,a_i):&=\sum_{\bm{a}_{-i}}\bm\mu^{\theta_{-i}}(\bm{a}_{-i}|s)r(s,a_i,\bm{a}_{-i}),    \\
            \overline{f_i^\theta}(s,a_i):&=\sum_{\bm{a}_{-i}}\bm\mu^{\theta_{-i}}(\bm{a}_{-i}|s)f^{\theta}(s,a_i,\bm{a}_{-i}).
        \end{align*}
        According to these definitions above, we have
        \begin{align*}
            J_{\epsilon}&=\sum\limits_{s,a}\pi^{\theta_\epsilon}(s)\theta_{\epsilon}(\bm{a}|s)f^{\theta_{\epsilon}}(s,\bm{a}) = \sum\limits_{s,a_i}\pi^{\theta_\epsilon}(s)\theta_{i,\epsilon}(a_i|s)\overline{f^{\theta_{\epsilon}}_i}(s,a_i),
        \end{align*}
        where $\overline{f^{\theta_{\epsilon}}_i}(s,a_i)= \sum\limits_{\bm{a}_{-i}} \theta_{-i}(\bm{a}_{-i}|s)f^{\theta_\epsilon}(s,a_i,\bm{a}_{-i})$. Next, we compute the first-order and second-order derivatives of $\overline{f^{\theta_{\epsilon}}_i}(s,a_i)$, respectively. For the first-order derivative, we have 
        \begin{align*}
            \frac{\text{d} \overline{f^{\theta_{\epsilon}}_i}(s,a_i)}{\text{d} \epsilon} = 2\beta (\overline{r_i}(s,a_i)-\eta^{\theta_\epsilon})\frac{\text{d} \eta^{\theta_\epsilon}}{\text{d} \epsilon},
        \end{align*}
        \begin{align*}
            \frac{\text{d}^2 \overline{f^{\theta_{\epsilon}}_i}(s,a_i)}{\text{d} \epsilon^2} = 2\beta \left[\left(\overline{r_i}(s,a_i)-\eta^{\theta_\epsilon}\right)\frac{\text{d}^2 \eta^{\theta_\epsilon}}{\text{d} \epsilon^2} - \left(\frac{\text{d} \eta^{\theta_\epsilon}}{\text{d} \epsilon} \right)^2 \right].
        \end{align*}
    
        For TSGs, it holds that $\nabla_{\bm\mu} J^{\bm\mu}=(\frac{\partial J^{\bm\mu}}{\partial \mu_1}, \dots, \frac{\partial J^{\bm\mu}}{\partial \mu_N})^T$. To investigate the smooth property of the mean-variance performance function, we have
        \begin{align*}
            \left\| \nabla_{\bm\mu} J^{\bm\mu} - \nabla_{\bm\mu} J^{\bm\mu'} \right\|_2^2 &= \sum\limits_{i=1}^N \left\| \nabla_{\mu_i} J^{\bm\mu} - \nabla_{\mu_i} J^{\bm\mu'} \right\|_2^2  \notag\\
            & \le \sum\limits_{i=1}^N \sum\limits_{j=1}^N \left\| \nabla_{\mu_i} J^{(\mu_{1\sim j-1}, \mu'_{j \sim N})} - \nabla_{\mu_i} J^{(\mu_{1\sim j}, \mu'_{j+1 \sim N})} \right\|_2^2,
        \end{align*}
        where $J^{\bm\mu'}=J^{(\mu_{1\sim 0}, \mu'_{1 \sim N})}$, $J^{\bm\mu}=J^{(\mu_{1\sim N}, \mu'_{N+1 \sim N})}$.
    
        \cite{cheng2024provable} show  that  $\left\| \frac{\text{d} \pi^{\theta_\epsilon}}{\text{d} \epsilon} \right\|_2 \le \frac{\kappa_0 \sqrt{SA_{\text{max}}}}{2}$, $\left\| \frac{\text{d}^2 \pi^{\theta_\epsilon}}{\text{d} \epsilon^2} \right\|_2 \le \kappa_0^2 S A_{\text{max}}$, and for the average performance function  $\eta$, it holds that
        \begin{align*}
            \frac{\text{d} \eta^{\theta_\epsilon}}{\text{d} \epsilon}= \sum\limits_{s,a_i} \pi^{\mu_\epsilon}(s) u_{s,a_i} \overline{r_i}(s,a_i) + \sum\limits_{s,a_i} \frac{\text{d} \pi^{\mu_\epsilon}(s)}{\text{d} \epsilon} \theta_{i,\epsilon}(a_i|s) \overline{r_i}(s,a_i),
        \end{align*}
        
        \begin{align*}
            \left| \frac{\text{d} \eta^{\theta_\epsilon}}{\text{d} \epsilon} \right| 
            &\le \left\|\pi^{\theta_\epsilon} \right\|_2 \sqrt{\sum\limits_s (\sum\limits_{a_i} u_{s,a_i}\overline{r_i}(s,a_i)})^2+ \left\|\frac{\text{d} \pi^{\theta_\epsilon}}{\text{d} \epsilon} \right\|_1  \le \sqrt{A_{\text{max}}}(1+ \frac{\kappa_0 S}{2}),
        \end{align*}

        Next, we first derive the first-order derivative of the mean-variance performance function with respect to $\epsilon$,
        \begin{align*}
            \frac{\text{d} J_\epsilon}{\text{d} \epsilon}=&\underbrace
            {\sum\limits_{s,a_i}\pi^{\theta_\epsilon}(s) u_{s,a_i} \overline{f^{\theta_{\epsilon}}_i}(s,a_i)}_{\text{Part A}} \notag\\
            &+ \underbrace
            {\sum\limits_{s,a_i} \frac{\text{d} \pi^{\theta_\epsilon}(s)}{\text{d} \epsilon} \theta_{i,\epsilon}(a_i|s) \overline{f^{\theta_{\epsilon}}_i}(s,a_i)}_{\text{Part B}} \notag\\
            &+ \underbrace
            {\sum\limits_{s,a_i}  \pi^{\theta_\epsilon}(s) \theta_{i,\epsilon}(a_i|s) \frac{\text{d} \overline{f^{\theta_{\epsilon}}_i}(s,a_i)}{\text{d} \epsilon}}_{\text{Part C}}.
        \end{align*}
    
        To compute the second-order derivative, we take the derivative of each of the three parts in the above expression separately. The derivative of Part A is as follows:
        \begin{equation*}
            \frac{\text{d} \text{Part A}}{\text{d} \epsilon} = \sum\limits_{s,a_i} \frac{\text{d} \pi^{\theta_\epsilon}(s)}{\text{d} \epsilon} u_{s,a_i} \overline{f^{\theta_{\epsilon}}_i}(s,a_i) + \sum\limits_{s,a_i} \pi^{\theta_\epsilon}(s) u_{s,a_i} \frac{\text{d} \overline{f^{\theta_{\epsilon}}_i}(s,a_i)}{\text{d} \epsilon}.
        \end{equation*}
        The derivative of Part B is as follows:
        \begin{align*}
            \frac{\text{d} \text{Part B}}{\text{d} \epsilon} 
            & = \sum\limits_{s,a_i} \frac{\text{d} \pi^{\theta_\epsilon}(s)}{\text{d} \epsilon} u_{s,a_i} \overline{f^{\theta_{\epsilon}}_i}(s,a_i) \notag\\
            & + \sum\limits_{s,a_i} \frac{\text{d} \pi^{\theta_\epsilon}(s)}{\text{d} \epsilon} \theta_{i,\epsilon}(s,a_i) \frac{\text{d} \overline{f^{\theta_{\epsilon}}_i}(s,a_i)}{\text{d} \epsilon} \notag\\
            & + \sum\limits_{s,a_i} \frac{\text{d}^2 \pi^{\theta_\epsilon}(s)}{\text{d} \epsilon^2} \theta_{i,\epsilon}(s,a_i) f^{\theta_{\epsilon}}_i.
        \end{align*}
        The derivative of Part C is as follows:
        \begin{align*}
            \frac{\text{d} \text{Part C}}{\text{d} \epsilon} 
            & = \sum\limits_{s,a_i} \frac{\text{d} \pi^{\theta_\epsilon}(s)}{\text{d} \epsilon} \theta_{i,\epsilon}(s,a_i) \frac{\text{d}\overline{f^{\theta_{\epsilon}}_i}}{\text{d}\epsilon} \notag\\
            & + \sum\limits_{s,a_i} \pi^{\theta_\epsilon}(s) u_{s,a_i} \frac{\text{d} \overline{f^{\theta_{\epsilon}}_i}(s,a_i)}{\text{d} \epsilon} \notag\\
            & + \sum\limits_{s,a_i} \pi^{\theta_\epsilon}(s) \theta_{i,\epsilon}(s,a_i) \frac{\text{d}^2 \overline{f^{\theta_{\epsilon}}_i}(s,a_i)}{\text{d} \epsilon^2}.
        \end{align*}
        By arranging the results, we obtain that the second-order derivative of the mean-variance performance function with respect to $\epsilon$ is:
        \begin{align*}
            \frac{\text{d}^2 J}{\text{d} \epsilon^2}=&
                \frac{\text{d} \text{Part A}}{\text{d} \epsilon} + \frac{\text{d} \text{Part B}}{\text{d} \epsilon} + \frac{\text{d} \text{Part C}}{\text{d} \epsilon} \notag\\
             &= \underbrace{
                 2 \sum\limits_{s,a_i} \frac{\text{d} \pi^{\theta_\epsilon}(s)}{\text{d} \epsilon} u_{s,a_i} \overline{f^{\theta_{\epsilon}}_i}(s,a_i) 
                }_{\textcircled{1}} \notag\\
             & +  \underbrace{
                 2 \sum\limits_{s,a_i} \pi^{\theta_\epsilon}(s) u_{s,a_i} \frac{\text{d} \overline{f^{\theta_{\epsilon}}_i}(s,a_i)}{\text{d} \epsilon}
              }_{\textcircled{2}} \notag\\
             & + \underbrace{
                 2 \sum\limits_{s,a_i} \frac{\text{d} \pi^{\theta_\epsilon}(s)}{\text{d} \epsilon} \theta_{i,\epsilon}(s,a_i) \frac{\text{d} \overline{f^{\theta_{\epsilon}}_i}(s,a_i)}{\text{d} \epsilon}
              }_{\textcircled{3}} \notag\\
             & + \underbrace{
                 \sum\limits_{s,a_i} \frac{\text{d}^2 \pi^{\theta_\epsilon}(s)}{\text{d} \epsilon^2} \theta_{i,\epsilon}(s,a_i) \overline{f^{\theta_{\epsilon}}_i}(s,a_i)
             }_{\textcircled{4}} \notag\\
             & + \underbrace{
                 \sum\limits_{s,a_i} \pi^{\theta_\epsilon}(s) \theta_{i,\epsilon}(s,a_i) \frac{\text{d}^2 \overline{f^{\theta_{\epsilon}}_i}(s,a_i)}{\text{d} \epsilon^2}
             }_{\textcircled{5}}.
        \end{align*}
    
        Based on the above result, we derive an upper bound for each of the five terms separately. For the first term, we have:
        \begin{align*}
            | \textcircled{1} | &\le  2 \left \|\frac{\text{d} \pi^{\theta_\epsilon}}{\text{d} \epsilon} \right \|_2 \sqrt{\sum\limits_s(\sum\limits_{a_i} u_{s,a_i} \overline{f^{\theta_{\epsilon}}_i}(s,a_i))^2} \notag\\
            & \le 2 \left \|\frac{\text{d} \pi^{\theta_\epsilon}}{\text{d} \epsilon} \right \|_2 \sqrt{A_i} f_{\text{max}} \left \| u_i \right \|_2 \notag\\
            & \le \kappa_0 f_{\text{max}} A_{\text{max}} \sqrt{S}.
        \end{align*}
        where $f_{\text{max}}$ represents the upper bound of the magnitude of the surrogate reward function $f$.
        For the second term, we have:
        \begin{align*}
            \left|\textcircled{2}\right| &= \left|4\beta \sum\limits_{s,a_i} \pi^{\theta_\epsilon} u_{s,a_i} (\overline{r_i}(s,a_i)-\eta^{\theta_\epsilon})\frac{\text{d} \eta^{\theta_{\epsilon}}}{\text{d} \epsilon}\right| \notag\\
            & \le 4\beta \left\| \pi^{\theta_\epsilon} \right\|_2 \sqrt{\sum\limits_s [\sum\limits_{a_i} u_{s,a_i}(\overline{r_i}(s,a_i)-\eta^{\theta_\epsilon})]^2} \left| \frac{\text{d} \eta^{\theta_{\epsilon}}}{\text{d} \epsilon}\right| \notag\\
            & \le 4\beta \cdot \sqrt{A_i} \cdot (\sqrt{A_i}+\frac{\kappa_0}{2}S \sqrt{A_i}) \notag\\
            & \le 4 \beta A_{\text{max}}(1+ \frac{\kappa_0}{2} S ).
        \end{align*}
        For the third term, we have:
        \begin{align*}
            \left|\textcircled{3}\right| &=\left| 4 \beta \sum\limits_{s,a_i} \frac{\text{d} \pi^{\theta_\epsilon}}{\text{d} \epsilon} \theta_{i,\epsilon}(a_i|s) (\overline{r_i}(s,a_i)-\eta^{\theta_\epsilon})\frac{\text{d} \eta^{\theta_{\epsilon}}}{\text{d} \epsilon} \right| \notag\\
            & \le 4\beta \left\| \frac{\text{d}\pi^{\theta_\epsilon}}{\text{d}\epsilon} \right\|_1  \cdot \left| \frac{\text{d} \eta^{\theta_{\epsilon}}}{\text{d} \epsilon}\right| \notag\\
            & \le 4\beta \sqrt{S}\left\| \frac{\text{d}\pi^{\theta_\epsilon}}{\text{d}\epsilon} \right\|_2 \sqrt{A}(1+\frac{\kappa_0S}{2}) \notag\\
            & \le \beta\kappa_0 S A_{\text{max}} (2+\kappa_0 S).
        \end{align*}
        For the fourth term, we have:
        \begin{align*}
            \left|\textcircled{4}\right| 
            & \le \left\| \frac{\text{d}^2 \pi^{\theta_\epsilon}}{\text{d} \epsilon^2}  \right\|_1 \cdot f_{\text{max}} \notag\\
            & \le \kappa_0^2 S^{\frac{3}{2}} A_{\text{max}} f_{\text{max}}.
        \end{align*}
        For the fifth term, we have:
        \begin{align*}
            \left|\textcircled{5}\right| 
            & = \left|  \sum\limits_{s,a_i} \pi^{\theta_\epsilon} \theta_{i,\epsilon}(a_i|s) \cdot 2 \beta \left[(\overline{r_i}(s,a_i) - \eta^{\theta_{\epsilon}}) \frac{\text{d}^2 \eta^{\theta_{\epsilon}}}{\text{d} \epsilon^2} - (\frac{\text{d} \eta^{\theta_{\epsilon}}}{\text{d} \epsilon})^2\right]  \right| \notag\\
            & = \left| 2\beta \left[(\eta^{\theta_\epsilon}-\eta^{\theta_\epsilon})  \frac{\text{d}^2 \eta^{\theta_{\epsilon}}}{\text{d} \epsilon^2} - \sum\limits_{s,a_i} \pi^{\theta_\epsilon} \theta_{i,\epsilon}(a_i|s) (\frac{\text{d} \eta^{\theta_{\epsilon}}}{\text{d} \epsilon})^2 \right] \right|\notag \\
            & \le 2\beta \left| \frac{\text{d} \eta^{\theta_{\epsilon}}}{\text{d} \epsilon} \right|^2 \notag\\
            & \le 2\beta A_{\text{max}} R_{\text{max}}^2 (1+\frac{\kappa_0 S}{2})^2,
        \end{align*}
        where $R_{\text{max}}$ represents the upper bound of the magnitude of the reward function $r$.

        In this proof, we assume that the reward function is normalized, i.e., $r \in [0, 1]$, then we have $R_{\text{max}}=1$ and $f_{\text{max}}=1+ \beta$. Furthermore, we have
        \begin{align*}
            \left| \frac{\text{d}^2 J}{\text{d} \epsilon^2}\right| &\le |\textcircled{1}|  + |\textcircled{2}| + |\textcircled{3}|  + |\textcircled{4}| + |\textcircled{5}|  \notag\\
            & \le 6\beta A_{\text{max}}(1+\frac{\kappa_0}{2}S)^2 + (1+\beta)\kappa_0A_{\text{max}} \sqrt{S} (\kappa_0 S + 1) = L .
        \end{align*}

        In a similar manner, we consider the successive second-order differentiation of the performance function $J_{\epsilon,\tau}$ with respect to $\epsilon$ and $\tau$. As the derivation is similar to the above, it is omitted here. We obtain:
        \begin{align*}
            \frac{\partial^2 J_{\epsilon,\tau}}{\partial \epsilon \partial\tau} 
            \le 6\beta A_{\text{max}}(1+\frac{\kappa_0}{2}S)^2 + (1+\beta) A_{\text{max}} (\kappa_0 S+\kappa_0^2 S^{\frac{3}{2}} +\kappa_0 \sqrt{S} + 1) = \frac{L_{ J}}{N}.
        \end{align*}
        Then, the proof is finished.
\endproof
\subsection{Proof of Theorem~\ref{theorem:converge_rate}}
\label{proof:theorem_converge_rate}

\proof{}
Prior to the proof of Theorem~\ref{theorem:converge_rate}, we introduce the auxiliary Lemma~\ref{lemma:improve_lower_bound} and Lemma~\ref{lemma:improve_upper_bound}.
    \begin{lemma}[\cite{bubeck2015convex}, Lemma~3.6]
        \label{lemma:improve_lower_bound}
        Assume that $J(\theta)$ is $L_J$-smooth with respect to $\theta \in \mathcal{X}$. Define the gradient mapping as follows:
        \begin{align}
            G^\alpha(\theta):=\frac{1}{\alpha}\left(\text{Proj}_{\mathcal{X}}(\theta+\alpha\nabla J(\theta))-\theta\right). \notag
        \end{align}
        Let $\theta^+=\theta + \alpha G^\alpha(\theta) = \text{Proj}_{\mathcal{X}}(\theta+\alpha\nabla J(\theta))$, then for  $\alpha \le \frac{1}{L_J}$ we have
        \begin{align}
            J(\theta^+)-J(\theta)\ge\frac{\alpha}{2}\|G^\alpha(\theta)\|_2^2. \notag
        \end{align}
    \end{lemma}
    \begin{lemma}[\cite{agarwal2021theory}, Proposition~37]
        \label{lemma:improve_upper_bound}
        Assume that $J(\theta)$ is $L_J$-smooth over $\theta \in \mathcal{X}$. The projected gradient update is defined as $\theta^+ = \theta + \alpha G^\alpha(\theta)$. If $|G^\alpha(\theta)|_2 \leq \epsilon$, then:
        \begin{align}
            \max_{\theta \in \mathcal{X}} (\theta - \theta^+)^\top \nabla_\theta J(\theta^+)\le \epsilon(1+\alpha L_J). \notag
        \end{align}        
        % \begin{align}
        %     \max_{\theta+\delta\in\Delta(\mathcal{A})^S,\left\|\delta\right\|_2\leq1}\delta^\top\nabla_\mu J(\mu^{\prime})\le \epsilon(1+\alpha L_J).
        % \end{align}
        In particular, when $\alpha \leq \frac{1}{L_J}$, we have
        \begin{align}
            \max_{\theta \in \mathcal{X}} (\theta - \theta^+)^\top \nabla_\theta J(\theta^+)\le 2\epsilon.  \notag
        \end{align}        
    \end{lemma}

    Next, we prove Theorem~\ref{theorem:converge_rate}. According to Lemma~\ref{lemma:improve_lower_bound}, when $\alpha \leq \frac{1}{L_J}$, we have
    \begin{align}
        \frac{1}{K}\sum\limits_{k=0}^{K-1}\| G^\alpha(\theta)\|_2^2 
        &\le \frac{2 L_J}{K}\sum\limits_{k=0}^{K-1}\Big(J(\theta^{(k+1)})-J(\theta^{(k)})\Big) \notag\\
        & = \frac{2 L_J}{K}(J(\theta^{(K)}))-J(\theta^{(0)})) \notag\\
        & \le \frac{2 L_J}{K}(J_{\text{max}}-J_{\text{min}}). \notag
    \end{align}
    Let $\frac{2 L_J}{K}(J_{\text{max}} - J_{\text{min}}) = \frac{\epsilon}{2}$, which implies that $K = \frac{4 L_J (J_{\max} - J_{\min})}{\epsilon}$, then we have
    \begin{align}
        \min\limits_k \|G^\alpha(\theta^{(k)})\|\le \frac{\epsilon}{2}.  \notag
    \end{align}
    Let $k^* = \arg\min_k |G(\theta^{(k)})|$. By Lemma~\ref{lemma:improve_upper_bound}, we can conclude that $\theta^{(k+1)}$ is $\epsilon$-stationary. This completes the proof.

\endproof

\end{document}